\documentclass[times,twocolumn,final]{elsarticle}

\usepackage{Clean/medima}
\usepackage{framed,multirow}
\usepackage{amssymb}
\usepackage{latexsym}
\usepackage{threeparttable}
\usepackage{algorithmicx}
\usepackage{algorithm}
\usepackage{algpseudocode}

\usepackage{url}
\usepackage{appendix}
\usepackage{xcolor}
\usepackage{bbm}
\usepackage[colorlinks]{hyperref}
\usepackage{booktabs}
\definecolor{newcolor}{rgb}{.8,.349,.1}
\journal{Medical Image Analysis}

\usepackage{graphicx}
\usepackage{subfigure}
\usepackage{float}
\usepackage{bbding}
\usepackage{amsmath,amsfonts}
\usepackage{amsthm}
\usepackage{amssymb,amsopn}
\usepackage{bm} 

\newcommand{\vct}[1]{\boldsymbol{#1}} 
\usepackage{xspace}
\newcommand{\eg}{{e.g.}\xspace}

\usepackage[font=small,labelfont=bf,labelsep=period]{caption}  
\usepackage{longtable}
\newlength\savewidth\newcommand\shline{\noalign{\global\savewidth\arrayrulewidth
  \global\arrayrulewidth 1.5pt}\hline\noalign{\global\arrayrulewidth\savewidth}}
\usepackage{Clean/collab}
\begin{document}
\verso{Gao \textit{et~al.}}

\newcommand{\etal}{\textit{et al}.\xspace}
\newcommand{\fun}[1]{\mathcal{#1}}
\newcommand{\setpar}{\vspace{5pt}}
\newcommand{\methodname}{NEED\xspace}
\newcommand{\methodproj}{SPDiff\xspace}
\newcommand{\methodimg}{DGDiff\xspace}

\begin{frontmatter}
\title{Noise-Inspired Diffusion Model for Generalizable Low-Dose CT Reconstruction}

\author[1]{Qi Gao}
\ead{qgao21@m.fudan.edu.cn}
\author[1]{Zhihao Chen}
\author[2]{Dong Zeng}
\author[3]{Junping Zhang}
\author[4]{Jianhua Ma}
\ead{jhma@smu.edu.cn}
\author[1,5,6,7]{Hongming Shan\corref{cor}}
\ead{hmshan@fudan.edu.cn}


\cortext[cor]{Corresponding author}

\address[1]{Institute of Science and Technology for Brain-inspired Intelligence, Fudan University, Shanghai 200433, China}
\address[2]{School of Biomedical Engineering, Southern Medical University, Guangzhou, Guangdong 510515, China}
\address[3]{School of Computer Science, Fudan University, Shanghai 200433, China}
\address[4]{School of Life Science and Technology, Xi’an Jiaotong University, Xi’an, Shaanxi 710049, China}
\address[5]{MOE Frontiers Center
for Brain Science, Fudan University, Shanghai 200433, China}
\address[6]{Key Laboratory of Computational Neuroscience and Brain-Inspired Intelligence (Ministry of Education), Fudan University, Shanghai 200433, China}
\address[7]{State Key Laboratory of Brain Function and Disorders, Fudan University, Shanghai 200433, China}

\received{9 Dec 2024}
\finalform{xx xx 2025}
\accepted{xx xx 2025}
\availableonline{xx xx 2025}
\communicated{S. Sarkar}
\begin{abstract}
The generalization of deep learning-based low-dose computed tomography (CT) reconstruction models to doses unseen in the training data is important and remains challenging.
Previous efforts heavily rely on paired data to improve the generalization performance and robustness through   
 collecting either diverse CT data for re-training or a few test data for fine-tuning.
Recently, diffusion models have shown promising and generalizable performance in low-dose CT (LDCT) reconstruction, however, 
they may produce unrealistic structures due to the CT image noise deviating from Gaussian distribution and imprecise prior information from the guidance of noisy LDCT images.
In this paper, we propose a \underline{n}ois\underline{e}-inspir\underline{e}d \underline{d}iffusion model for generalizable LDCT reconstruction, termed \methodname, which tailors diffusion models for noise characteristics of each domain.
First, we propose a novel shifted Poisson diffusion model to denoise projection data, which aligns the diffusion process with the noise model in pre-log LDCT projections.
Second, we devise a doubly guided diffusion model to refine reconstructed images, which leverages LDCT images and initial reconstructions to more accurately locate prior information and enhance reconstruction fidelity.
By cascading these two diffusion models for dual-domain reconstruction, 
our \methodname requires only normal-dose data for training and can be effectively extended to various unseen dose levels during testing via a time step matching strategy.
Extensive qualitative, quantitative, and segmentation-based evaluations on two datasets demonstrate that our \methodname consistently outperforms state-of-the-art methods in reconstruction and generalization performance. Source code is made available at \url{https://github.com/qgao21/NEED}.
\end{abstract}

\end{frontmatter}

\begin{figure}[!t]
\centerline{\includegraphics[width=1\linewidth]{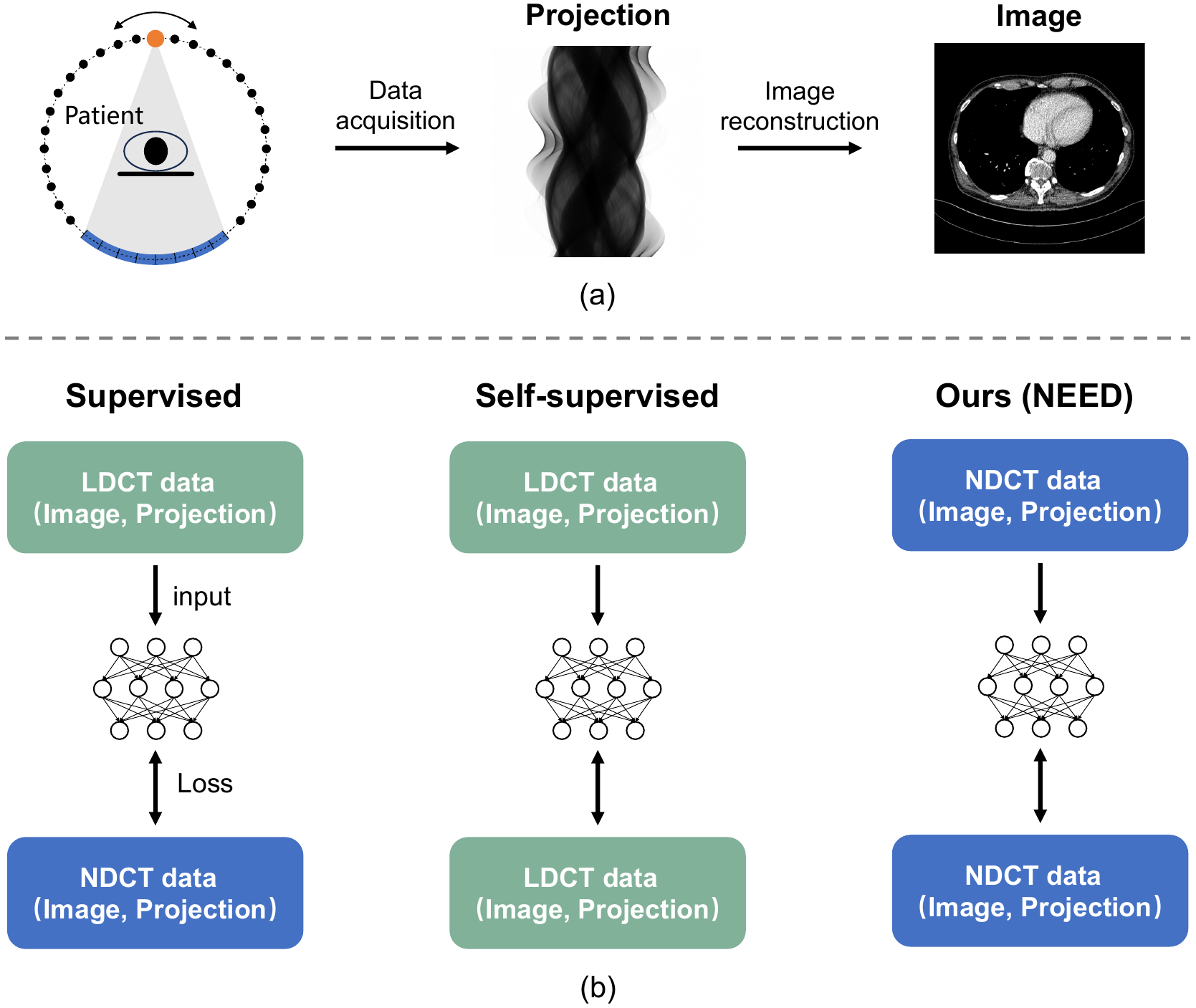}}
\caption{Overview of CT reconstruction and different learning-based reconstruction paradigms. (a) CT reconstruction pipeline. (b) Illustration of the differences in training strategies among supervised methods, self-supervised methods, and our \methodname.}
\label{fig:concept}
\end{figure}

\section{Introduction}
Computed tomography (CT) plays a crucial role in clinical disease diagnosis, physical examination screening, and surgical navigation. While CT has undeniably propelled medical advancements, research indicates that the potential health risks associated with X-ray exposure cannot be overlooked~\citep{smith2009radiation, sodickson2009recurrent}.
In clinical practice, reducing radiation dose often involves lowering the tube current. However, this reduction inevitably leads to increasing noise and artifacts in the reconstructed images. To mitigate these effects, various model-based iterative reconstruction algorithms and classical image post-processing techniques have been proposed to enhance the quality of low-dose CT (LDCT) reconstruction~\citep{wang2006penalized, ma2011low, xie2017robust}.

Deep learning (DL)-based low-dose CT reconstruction algorithms have been widely studied and have demonstrated encouraging performance~\citep{shan2019competitive, wang2020deep}. These algorithms can be divided into three categories: 1) projection domain-based algorithms; 2) image domain-based algorithms; and 3) dual domain-based algorithms.
Projection domain-based algorithms, such as  AttnRDN~\citep{ma2021sinogram} and SIST~\citep{yang2022low}, are designed to correct the noisy projection signal captured by detectors. 
Image domain-based algorithms directly denoise analytical reconstructed LDCT images, bypassing the secondary degradation linked to the reconstruction and thereby increasing flexibility. Representative works include RED-CNN~\citep{chen2017low}, CPCE~\citep{shan2018d}, DU-GAN~\citep{huang2021gan}, AIGAN~\citep{fu2023aigan} and AdvEM~\citep{sharma2024adversarial}.
Dual domain-based algorithms merge the strengths of both projection domain and image domain-based methods, yielding superior performance.
For example, Yin~\etal devised a domain-progressive three-dimensional residual CNN, comprising a projection domain subnetwork, a filtered backprojection layer, and an image domain subnetwork, tailored for LDCT imaging~\citep{yin2019domain, zhou2022dudodr}.
Ge~\etal enhanced low-dose CT image reconstruction by performing parallel optimization and cross-communication on projection and image domain data, culminating in the use of triple-cross attention mechanisms to fuse dual-domain information effectively~\citep{ge2022ddpnet}.
However, these methods are confined to handling test data with doses encountered during training, and their performance declines sharply when dose of the test data differs from that of the training data.

To tackle this challenge, some works utilize parameter-dependent framework~\citep{xia2021ct} or one-shot learning framework~\citep{gao2024corediff} to enhance the generalization performance of DL-based LDCT reconstruction algorithms across various test dose scenarios.
However, they involve either collecting a large amount of paired training data for supervised training to encompass diverse scenarios or a few paired test data for fine-tuning. The scarcity of normal-/low-dose CT paired data restricts the applicability of these algorithms in clinical practice. Moreover, existing self-supervised reconstruction methods utilized LDCT data for training, with no assurance of generalization performance for unseen test doses~\citep{niu2022noise, niu2022self}.

Recently, a growing number of studies have started to investigate the application of diffusion models for unsupervised image restoration, which necessitate only clean data for training~\citep{wang2023dr2, fei2023generative, garber2024image}.
For instance, denoising diffusion restoration models incorporate the inverse problem of image restoration as a condition within the sampling process of diffusion models~\citep{kawar2022denoising}. A variational inference objective is used to iteratively solve the posterior distribution associated with the inverse problem.
Dn-Dp integrates the prior information from the denoising diffusion probabilistic model with the multiple maximum a posterior framework and incorporates LDCT images into the sampling process for iterative optimization~\citep{liu2023diffusion}.
In general, these studies adopt one of two strategies: 1) employing the pretrained diffusion model as an iterative Gaussian denoiser, or 2) leveraging the diffusion model to model the distribution of clean data, thereby serving as a generative prior that guides the image restoration towards this distribution. However, the noise in the CT image domain is complex and does not follow a Gaussian distribution.
When noise and artifacts in LDCT images are severe, these models may generate unrealistic pixels or structures due to their inability to accurately locate the prior information, potentially affecting radiologists' diagnoses. Xie \etal redefine the diffusion process for different denoising tasks, enabling the diffusion model to handle not only Gaussian noise but also gamma and Poisson noise~\citep{xie2023diffusion}. Nevertheless, this framework demands noisy-clean image pairs for supervised learning. In addition, noise within both the CT projection and image domains does not conform to any of the above noise types.

In this paper, we propose a NoisE-inspirEd Diffusion model for low-dose CT reconstruction, termed \methodname, which is generalizable to unseen dose levels.
Considering the noise distribution of pre-log LDCT measurements, we design a novel shifted Poisson diffusion model for projection data denoising, using a forward process that progressively adds shifted Poisson noise to align with the noise model.
Due to the complex and unknown noise in the LDCT image domain, we incorporate a doubly guided diffusion model to characterize normal-dose CT (NDCT) image distribution as the generative prior for reconstructed image refinement. 
To improve the accuracy of prior information localization, this model uses LDCT images and reconstructed images to simultaneously modulate images before and after U-Net denoising during the sampling process.
Furthermore, we develop a time step matching strategy to control the degree of denoising and refinement, enabling it to generalize to LDCT data at unseen dose levels and accelerate sampling.
Notably, our method is trained with only pre-log NDCT projections and images, without access to any low-dose data. Fig.~\ref{fig:concept} illustrates the CT reconstruction pipeline and the training differences among supervised methods, self-supervised methods, and our \methodname. Both the supervised and self-supervised methods are trained for specific low-dose. In contrast, our \methodname relies solely on NDCT data, enabling generalizable reconstruction for multiple dose levels. 

The contributions of this work can be summarized as follows. 
First, we propose a novel model \methodname for generalizable LDCT reconstruction, which tailors diffusion models for noise characteristics of projection and image domains, only requires NDCT data for training, and is generalizable to unseen dose levels. 
Second, we introduce a shifted Poisson diffusion model to denoise LDCT projection data, which aligns the noise model to effectively mimic the degradation process of pre-log projection.
Third, we devise a doubly guided diffusion model to refine reconstructed images, modulating noisy and denoised images during the sampling process to enhance localization accuracy of prior information and data fidelity.
Fourth, we further develop a time step matching strategy that adaptively selects the optimal sampling steps for each test case, enabling generalization to various dose levels and accelerating inference.
Fifth, extensive experimental results on two datasets demonstrate the superior reconstruction and generalization performance of \methodname over state-of-the-art methods.
\begin{figure*}[!t]
\centerline{\includegraphics[width=1\linewidth]{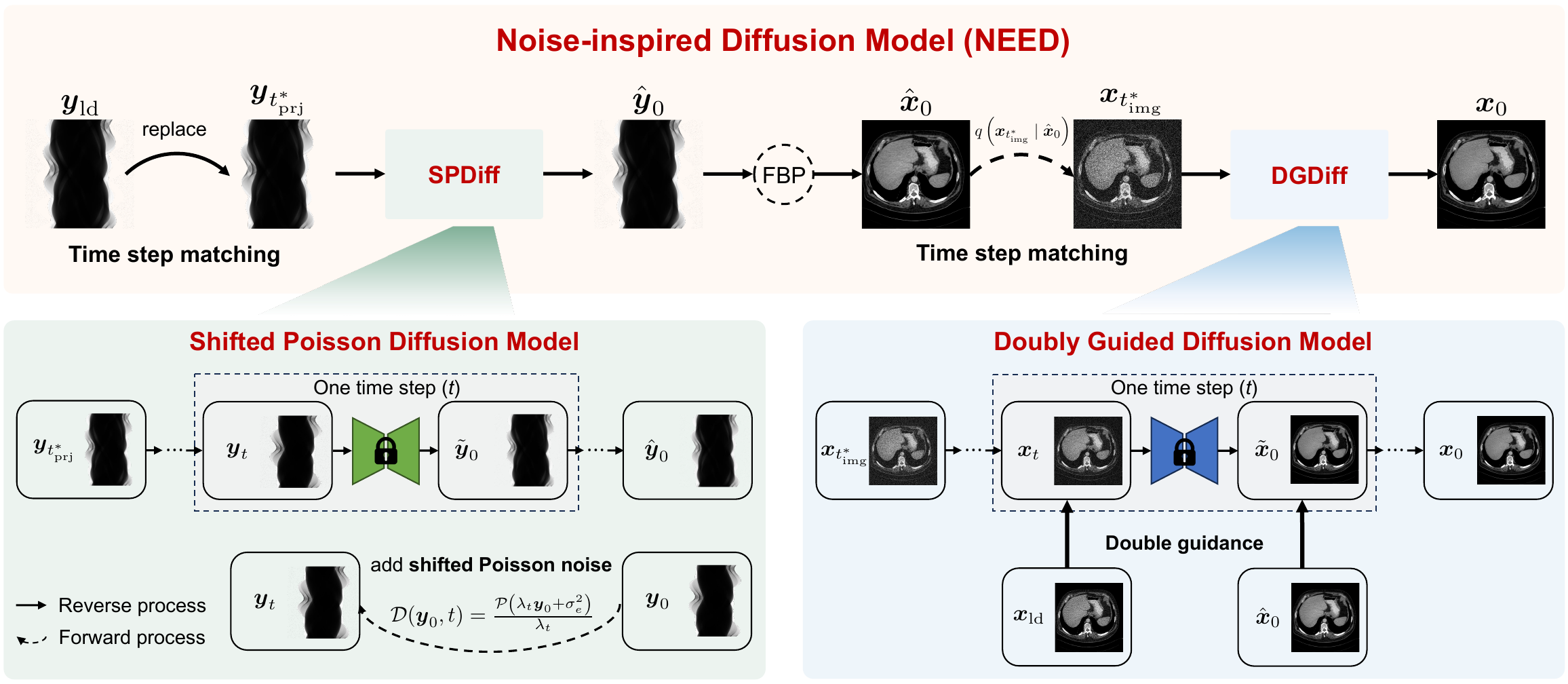}}
\caption{Overview of our \methodname for generalizable low-dose CT reconstruction. Our \methodname cascades two components: 1) a shifted Poisson diffusion model (\methodproj) for pre-log LDCT projection data denoising and 2) a doubly guided diffusion model (\methodimg) for reconstructed image refinement. Then we utilize a time step matching strategy to control the denoising and refinement while reducing sampling steps.}
\label{fig:overall_framework}
\end{figure*}

\section{Preliminaries}\label{sec:preliminaries}
\subsection{CT Noise Model}
Low-dose CT scans introduce noise into the pre-log projection data, which is then propagated through non-linear reconstruction processes, resulting in distinct statistical characteristics in both the projection and image domains. In pre-log LDCT raw measurements two major types of noise are: 1) quantum noise due to the X-ray photon fluctuations, and 2) electronic noise caused by the data acquisition system. Pre-log LDCT raw measurements can be mathematically expressed as a ``Poisson+Gaussian'' model~\citep{wang2017hybrid, fu2016comparison}:
\begin{align}
\vct{m}_\mathrm{ld} \sim \mathcal{P}\left(\overline{\vct{m}}(\vct{x})\right)+\mathcal{N}\left(0, \sigma_{e}^{2}\right),
\label{eq:Poisson_Gaussian_model}
\end{align}
where $\vct{m}_\mathrm{ld}$ is the pre-log LDCT raw measurement, $\overline{\vct{m}}(\vct{x})$ the corresponding expected measurement related to the image $\vct{x}$ of linear attenuation coefficient, $\mathcal{P}$ the Poisson distribution, $\mathcal{N}$ the Gaussian distribution, and $\sigma_{e}$ the standard deviation of electronic noise. The expectation $\overline{\vct{m}}(\vct{x})$ can be approximated by a clean NDCT measurement $\vct{m}_{0}$. In this work, we focus on the projection data normalized by the corresponding number of X-ray incident photons $I_\mathrm{ld}$ and $I_\mathrm{0}$, defined as:
\begin{align}
\vct{y}_\mathrm{ld}=\frac{\vct{m}_\mathrm{ld}}{I_\mathrm{ld}},\quad\vct{y}_\mathrm{0}=\frac{\vct{m}_\mathrm{0}}{I_0},
\label{eq:normalized_projection}
\end{align}
where $\vct{y}_\mathrm{ld}$ and $\vct{y}_{0}$ are normalized pre-log projections. In practical scanning scenarios, the number of X-ray incident photons $I_\mathrm{0}$ can be sourced from the data acquisition system or be estimated from either air calibration scans or unattenuated background regions within raw measurements, while $\sigma_{e}$ can be determined via the standard detector calibration procedure~\citep{hsieh2003computed, Thorsten2003computed, fu2016comparison}. In certain scenarios, only the post-log projection data (also known as sinograms) are accessible following calibration, preprocessing, and logarithmic transformation of the raw measurements.
These operations alter the noise distribution, making it challenging to develop a precise physics-based noise model for sinograms as could be done with pre-log raw measurements. Despite these challenges, phantom studies have demonstrated that the ``Poisson+Gaussian'' noise present in pre-log measurements can be effectively approximated by a signal-dependent Gaussian noise model in sinograms~\citep{li2004nonlinear, ma2012variance}.

In the image domain, log transformation, interpolation, and filtering during reconstruction alter noise distribution and introduce spatial correlation, making the noise model unknown. Some works approximate this complex noise using Gaussian mixture models or spatially correlated noise models~\citep{li2022noise, divel2019accurate}. Others characterize clean NDCT image properties, incorporating them as prior information in iterative reconstruction or deep learning algorithms.

\subsection{Diffusion Model}
Diffusion models are a class of explicit generative models designed with Markov chains. They progressively degrade the data distribution to a simpler known form through a forward (diffusion) process and then generate data iteratively from this distribution in a reverse (sampling) process. The Gaussian diffusion model is a prominent representative among them that constructs the forward process by progressively adding varying levels of Gaussian noise to the target data, ${\vct{x}_{0} \sim Q\left(\vct{x}_{0}\right)}$~\citep{ho2020denoising, kazerouni2023diffusion}:
\begin{align}
q\left(\vct{x}_{t}|\vct{x}_{0}\right)=\mathcal{N}\left(\vct{x}_{t} ; \sqrt{\bar{\alpha}_{t}} \vct{x}_{0},\left(1-\bar{\alpha}_{t}\right) \vct{I}\right),
\label{eq:Gaussian_diffusion_q_t}
\end{align}
where $\bar{\alpha}_{t}$ is the hyperparameter to control noise levels. The reverse process starts with a random Gaussian noise ${\vct{x}_{T} \sim \mathcal{N}(\mathbf{0}, \boldsymbol{I})}$ and generates clean data $\vct{x}_{0}$ through iterative denoising:
\begin{align}
p_{\vct{\theta}}\left(\vct{x}_{0:T}\right):=p\left(\vct{x}_{T}\right) \prod\nolimits_{t=1}^{T} p_{\vct{\theta}}\left(\vct{x}_{t-1} | \vct{x}_{t}\right),
\label{eq:Gaussian_diffusion_reverse_process}
\end{align}
where
\begin{align}
p_{\vct{\theta}}\left(\vct{x}_{t-1} | \vct{x}_{t}\right):=\mathcal{N}\left(\vct{x}_{t-1} ; \vct{\mu}_{\vct{\theta}}\left(\vct{x}_{t}, t\right), \vct{\Sigma}_{\vct{\theta}}\left(\vct{x}_{t}, t\right)\right).
\label{eq:Gaussian_diffusion_p_t_Markov}
\end{align}

Here, mean $\vct{\mu}_{\vct{\theta}}\left(\vct{x}_{t}, t\right)$ and variance $\vct{\Sigma}_{\vct{\theta}}\left(\vct{x}_{t}, t\right)$ can be estimated by a noise estimation network $\vct{\epsilon}_{\vct{\theta}}$ with parameter $\vct{\theta}$. During sampling process, the trained $\vct{\epsilon}_{\vct{\theta}}$ is used to estimate the denoised image $\tilde{\vct{x}}_{0}$ at each time step:
\begin{align}
\tilde{\vct{x}}_{0} = \frac{{\vct{x}_{t}}-{\sqrt{1-\bar{\alpha}_{t}} \vct{\epsilon}_{\vct{\theta}}\left(\vct{x}_{t}, t\right)}}{\sqrt{\bar{\alpha}_{t}}}.
\label{eq:Gaussian_diffusion_denoised_x0}
\end{align}

Although the Gaussian diffusion model has demonstrated impressive performance in the image generation field, its fixed forward process limits its utility in other tasks such as image restoration.
Cold diffusion breaks free from the theoretical limitations of the classical Gaussian diffusion models, providing greater flexibility~\citep{bansal2022cold}. It allows various types of degradation, \eg adding noise, blurring, and downsampling, with the forward process being formulated as a custom degradation operator:
\begin{align}
\vct{x}_{t}=\fun{D}(\vct{x}_{0}, t).
\label{eq:commonly_used_degradation_operator}
\end{align}

Subsequently, a ``restoration-redegradation'' sampling process is developed to generate the image from a simple distribution. This process relies on a restoration network, $\fun{R}_{\phi}$, which is parameterized by $\phi$ and trained to predict $\vct{x}_{0}$ from $\vct{x}_{t}$ and $t$:
\begin{align}
\tilde{\vct{x}}_{0} = \fun{R}_{\phi}(\vct{x}_{t}, t) \approx \vct{x}_{0}.
\label{eq:cold_diff_restoration_operator}
\end{align}
\section{Methodology}
\label{sec:method}
In this section, we first present the shifted Poisson diffusion model (\methodproj) for pre-log LDCT projection data denoising. Then, we introduce a doubly guided diffusion model (\methodimg) as the generative prior for reconstructed image refinement. Finally, we highlight the dual-domain diffusion model \methodname, which integrates both components into a cascade pipeline for generalizable LDCT reconstruction, leveraging a time step matching strategy for generalization to unseen dose levels. Fig.~\ref{fig:overall_framework} presents the overall framework of our \methodname.

\subsection{Shifted Poisson Diffusion Model for Projection Domain}\label{sec:shifted_poisson_diffusion_model}
Previous works have shown that the pretrained diffusion models can directly function as denoisers, yet their capability is restricted to removing additive Gaussian noise due to the inherent Gaussian diffusion process. 
To adapt the diffusion model for pre-log LDCT projection data denoising, in this work, we incorporate the noise model into the cold diffusion model to construct a tailored diffusion process.
Since Poisson noise is the dominant source of degradation and is challenging to handle, we employ the shifted Poisson model as a more tractable alternative to Eqs.~\eqref{eq:Poisson_Gaussian_model} and~\eqref{eq:normalized_projection}, defining the degradation operator as:
\begin{align}
\fun{D}(\vct{y}_{0}, t)=\mathcal{P}\left(\lambda_{t} \vct{y}_{0}+\sigma_{e}^{2}\right) / \lambda_{t},\quad1\leq{t}\leq{\tau},
\label{eq:shifted_Poisson_diffusion_q_t}
\end{align}
where $\lambda_{t}$ is a time-dependent hyperparameter controlling the intensity of shifted Poisson noise and $\lambda_{1}>\lambda_{2}>\cdots>\lambda_{\tau}$; here, $\tau$ is the total diffusion time steps of the shifted Poisson diffusion model. This degradation operator enables an approximate matching of the degraded projection from the diffusion process to the pre-log LDCT projection due to the alignment of the noise model.

The sampling process starts from the pre-log LDCT projection and iteratively removes the shifted Poisson noise. To reduce accumulated errors, we utilize an improved sampling algorithm instead of the original ``restoration-redegradation'' one, predicting the projection at time step $t-1$:
\begin{align}
\vct{y}_{t-1}=\frac{\lambda_{t} \vct{y}_{t}+\mathcal{P}\left(\left(\lambda_{t-1}-\lambda_{t}\right) \tilde{\vct{y}}_{0}\right)}{\lambda_{t-1}},
\label{eq:shifted_Poisson_diffusion_sampling}
\end{align}
where $\tilde{\vct{y}}_{0}$ is estimated by the restoration network $\fun{R}_{\vct{\phi}}\left(\vct{y}_{t}, t\right)$. The parameter $\vct{\phi}$ can be optimized by minimizing following objective function:
\begin{align}
\min _{\vct{\phi}} \mathbb{E}_{\vct{y}_{0}, t} \left\|\vct{y}_{0}-\fun{R}_{\vct{\phi}}\left(\vct{y}_{t}, t\right)\right\|^{2},
\label{eq:shifted_Poisson_diffusion_loss}
\end{align}
where $\vct{y}_{t}=\fun{D}(\vct{y}_{0}, t)$. We emphasize that the training of our shifted Poisson diffusion model solely utilizes $\vct{y}_{0}$, which is sampled from the distribution of pre-log NDCT projection data. 
Finally, the denoised projection, $\hat{\vct{y}}_{0}$, can be reconstructed into an image $\hat{\vct{x}}_{0}$ using filtered back-projection (FBP):
\begin{align}
\hat{\vct{x}}_{0}=\operatorname{FBP}(-\log(\hat{\vct{y}}_{0})).
\label{eq:shifted_Poisson_diffusion_FBP}
\end{align}

\subsection{Doubly Guided Diffusion Model for Image Domain}\label{sec:Image domain refinement}
Denoising in the projection domain reduces much noise but may introduce blurring and secondary artifacts into the initial reconstruction $\hat{\vct{x}}_{0}$, necessitating subsequent image domain refinement. However, the absence of a known noise model in the image domain makes it challenging to construct a customized diffusion model. An effective approach involves utilizing the Gaussian diffusion model to model the distribution of clean NDCT images, employing it as a prior to solve the image restoration inverse problem~\citep{yu2023unsupervised, liu2023diffusion}. Unlike \methodproj that functions as a specialized shifted Poisson denoiser, the pixel-independent Gaussian noise added within this diffusion model serves a distinct purpose. It is used to train the model to predict the score function (the gradient of the log-probability density), which represents the direction of change in the data distribution~\citep{song2020score}. Nevertheless, each LDCT image corresponds to multiple clean NDCT images, making it challenging to accurately locate the corresponding prior information when directly using degraded LDCT images to quantify the data fidelity.

To address this challenge, we design a doubly guided diffusion model that utilizes a pretrained Gaussian diffusion model as a generative prior and incorporates double guidance at each step of the sampling process: 1) employing the LDCT image $\vct{x}_\mathrm{ld}$ to guide the noisy image $\vct{x}_{t}$, and 2) leveraging the initial reconstruction $\hat{\vct{x}}_{0}$ to modulate the denoised image $\tilde{\vct{x}}_{0}$. 
Different from other diffusion prior-based works~\citep{fei2023generative, liu2023diffusion, wang2023dr2}, we use the reconstructed image $\hat{\vct{x}}_{0}$ with less noise rather than degraded image to guide $\tilde{\vct{x}}_{0}$, thereby avoiding the excessive introduction of noise and artifacts into the denoised image $\tilde{\vct{x}}_{0}$. Given the potential for information loss during the pre-log projection data denoising, we also introduce the LDCT image $\vct{x}_\mathrm{ld}$ as guidance in the sampling process. Specifically, we utilize $\vct{x}_\mathrm{ld}$ to guide the noisy image $\vct{x}_{t}$ before it is input into the noise estimation network $\vct{\epsilon}_{\vct{\theta}}$ to enhance data fidelity. The proposed \methodimg aims to diminish the degradation impact of $\vct{x}_\mathrm{ld}$ while preserving more original information.
The improved refinement pipeline can be expressed as follows~\citep{fei2023generative, sohl2015deep, dhariwal2021diffusion}:
\begin{align}
&\log p_{\vct{\theta}}\left(\vct{x}_{t-1} | \vct{x}_{t}, \vct{x}_\mathrm{ld}, \hat{\vct{x}}_{0}\right) \notag \\
=&\log \left(p_{\vct{\theta}}\left(\vct{x}_{t-1} | \vct{x}_{t}\right)p\left(\vct{x}_\mathrm{ld} | \vct{x}_{t}\right)p\left(\hat{\vct{x}}_{0} | \vct{x}_{t}\right)\right)+C,
\label{eq:image_domain_finetuning_pipeline}
\end{align}
where $C$ is a constant. $p\left(\vct{x}_\mathrm{ld} | \vct{x}_{t}\right)$ and $p\left(\hat{\vct{x}}_0 | \vct{x}_{t}\right)$ can be regarded as the probability that $\vct{x}_{t}$ will be denoised to $\tilde{\vct{x}}_0$ consistent to $\vct{x}_\mathrm{g}$. $\vct{x}_\mathrm{g}$ represents the guidance image, which in our method corresponds to both $\vct{x}_0$ and $\vct{x}_\mathrm{ld}$.
Since the closed form of $p\left(\vct{x}_\mathrm{g} | \vct{x}_{t}\right)$ is not tractable, an effective approximation involving using a distance function $L(\cdot)$ to constrain the consistency between $\vct{x}_t$ and $\vct{x}_\mathrm{g}$~\citep{zhu2023denoising, fei2023generative}. We utilize the pixel-level $L_1$ distance as $L(\cdot)$, and the optimized $\tilde{\vct{x}}_{t}$ from $p\left(\vct{x}_\mathrm{ld} | \vct{x}_{t}\right)$ can be calculated as follows:
\begin{align}
\tilde{\vct{x}}_{t}=s_{1}\vct{x}_{t} + (1-s_{1})(\sqrt{\bar{\alpha}_{t}} \vct{x}_\mathrm{ld}),
\label{eq:image_domain_finetuning_guidance_xt}
\end{align}
where $\sqrt{\bar{\alpha}_{t}}$ represents a scaling coefficient to ensure that the mean of $\vct{x}_\mathrm{ld}$ aligns with that of $\vct{x}_{t}$.

For $p\left(\hat{\vct{x}}_0 | \vct{x}_{t}\right)$, we can also process it by adding Gaussian noise at the same level as $\vct{x}_{t}$ to the initial reconstruction $\hat{\vct{x}}_{0}$ and employing a formulation similar to Eq.~\eqref{eq:image_domain_finetuning_guidance_xt}.
However, merely modulating the network input at each sampling step lacks optimization of the output image $\tilde{\vct{x}}_{0}$. Considering that the quality of $\tilde{\vct{x}}_{0}$ directly correlates with $\vct{x}_{t}$ and our objective is to obtain an improved final reconstructed image $\vct{x}_{0}$, we employ a doubly guidance to optimize the $L_1$ distance between the initial reconstruction $\hat{\vct{x}}_0$ and the denoised image $\tilde{\vct{x}}_0$ rather than $\vct{x}_{t}$, achieving better information utilization. The final reconstructed image $\vct{x}_0$ at each sampling step can be defined as:
\begin{align}
\vct{x}_{0}=s_{2}\tilde{\vct{x}}_{0} + (1-s_{2})\hat{\vct{x}}_{0},
\label{eq:image_domain_finetuning_guidance_x0}
\end{align}
where the weighting parameters $s_{1}$ and $s_{2}$ are used to balance the trade-off between data fidelity and noise suppression.

\subsection{Inference Pipeline for Generalizable Reconstruction}\label{sec:dualdiff_for_dual_domain_denoising_generalization}
As shown in Fig.~\ref{fig:overall_framework}, our \methodname cascades the trained {\methodproj and \methodimg} for LDCT dual-domain reconstruction. Directly applying the full sampling process to each test LDCT data would be suboptimal due to the potential over-suppression of noise and redundant computational expense. Therefore, we design a time step matching strategy that adaptively selects an appropriate starting point to handle test data at various dose levels for generalizable reconstruction.
First, given a test pre-log LDCT projection $\vct{y}_\mathrm{ld}$ and its corresponding number of X-ray incident photons $I_\mathrm{ld}$, we find an appropriate $t_\mathrm{prj}^{*}$ where $\lambda_{{t}_\mathrm{prj}^{*}}$ is closest to $I_\mathrm{ld}$:
\begin{align}
t_\mathrm{prj}^{*} = \arg\min_{t}\left|\lambda_{t}-I_\mathrm{ld}\right|.
\label{eq:shifted_Poisson_diffusion_best_t1}
\end{align}

The sampling process initiates from time step $t_\mathrm{prj}^{*}$, utilizing $\vct{y}_\mathrm{ld}$ as the starting point. Since the noise level in $\vct{y}_\mathrm{ld}$ is primarily controlled by $I_\mathrm{ld}$, and the denoising level of the \methodproj is determined by $\lambda_t$, this time step matching strategy enhances the model's capacity to generalize to unseen dose levels in the projection domain. We employ the sampling algorithm in Eq.~\eqref{eq:shifted_Poisson_diffusion_sampling} and the reconstruction algorithm in Eq.~\eqref{eq:shifted_Poisson_diffusion_FBP} to obtain the initial reconstruction $\hat{\vct{x}}_{0}$.

Second, similar to the \methodproj, we perform the time step matching strategy based on $\sigma_\mathrm{ld}$ to determine the refinement steps required for $\hat{\vct{x}}_{0}$ within the \methodimg:
\begin{align}
t_\mathrm{img}^{*}=\arg\min\limits_{t}\left|\sqrt{1-\bar{\alpha}_{t}}-\hat{\sigma}_\mathrm{ld}\right|,
\label{eq:image_domain_finetuning_best_t2}
\end{align}
where $\sigma_\mathrm{ld}$ is the standard deviation between $\vct{x}_\mathrm{ld}$ and $\hat{\vct{x}}_{0}$: ${\sigma}_\mathrm{ld}=\operatorname{std}(\vct{x}_\mathrm{ld}-\hat{\vct{x}}_{0})$. Eq.~\eqref{eq:image_domain_finetuning_best_t2} indicates that  $\hat{\vct{x}}_{0}$ denoised by a pre-log projection with a lower dose requires more refinement steps. Considering that diffusion models focus on the low-frequency information of the generated image during a large number of early sampling steps, which is well restored in $\hat{\vct{x}}_0$, we can use only a small number of later time steps to compensate high-frequency details, thereby avoiding the generation of unrealistic structure and reducing the steps of refinement. After $t_\mathrm{img}^{*}$ steps of refinement, we obtain the final reconstruction image $\vct{x}_{0}$. The total sampling steps of our \methodname is $t_\mathrm{prj}^{*}+t_\mathrm{img}^{*}$ and the sampling (inference) procedure is shown in Alg.~\ref{alg:sampling}.

\begin{algorithm}[!t]
\caption{Inference for \methodname}
\label{alg:sampling}
\begin{algorithmic}[1]
\Require A test pre-log LDCT projection $\vct{y}_\mathrm{ld}$ and corresponding image $\vct{x}_\mathrm{ld}$, number of X-ray incident photons $I_\mathrm{ld}$
\Ensure Denoised image $\vct{x}_{0}$
\State \textbf{Load} the trained ${\fun{R}}_{\vct{\phi}}$ and ${\vct{\epsilon}}_{\vct{\theta}}$
\State \textbf{Calculate} ${t}_\mathrm{prj}^{*}$ by Eq.~\eqref{eq:shifted_Poisson_diffusion_best_t1}
\State \textbf{Replace} $\vct{y}_{{t}_\mathrm{prj}^{*}}$ by $\vct{y}_\mathrm{ld}$
\For{${t={t}_\mathrm{prj}^{*}, {t}_\mathrm{prj}^{*}-1, \ldots, 2}$}
\State $\tilde{\vct{y}}_{0} \leftarrow \fun{R}_{\vct{\phi}}\left(\vct{y}_{t}, t\right)$
\State \textbf{Calculate} $\vct{y}_{t-1}$ by Eq.~\eqref{eq:shifted_Poisson_diffusion_sampling}
\EndFor
\State $\hat{\vct{y}}_{0} \leftarrow \fun{R}_{\vct{\phi}}\left(\vct{y}_{1}, 1\right)$
\State \textbf{Reconstruct} $\hat{\vct{x}}_{0}$ by Eq.~\eqref{eq:shifted_Poisson_diffusion_FBP}
\State \textbf{Calculate} ${t}_\mathrm{img}^{*}$ and $\vct{x}_{{t}_\mathrm{img}^{*}}$ by Eq.~\eqref{eq:image_domain_finetuning_best_t2} and Eq.~\eqref{eq:Gaussian_diffusion_q_t}
\For{${t={t}_\mathrm{img}^{*}, {t}_\mathrm{img}^{*}-1, \ldots, 1}$}
\State \textbf{Modulate} $\vct{x}_{t}$ by Eq.~\eqref{eq:image_domain_finetuning_guidance_xt}
\State \textbf{Calculate} $\tilde{\vct{x}}_{0}$ by Eq.~\eqref{eq:Gaussian_diffusion_denoised_x0}
\State \textbf{Modulate} $\tilde{\vct{x}}_{0}$ by Eq.~\eqref{eq:image_domain_finetuning_guidance_x0}
\State \textbf{Calculate} $\vct{x}_{t-1}$ by Eq.~\eqref{eq:Gaussian_diffusion_q_t}
\EndFor
\end{algorithmic}
\end{algorithm}
\section{Experiment Results}
\label{sec:results}

\subsection{Datasets}
\noindent\textbf{Mayo 2016 dataset}\quad We evaluate the proposed \methodname on the publicly available CT dataset \textit{2016 NIH-AAPM-Mayo Clinic Low Dose CT Grand Challenge}~\citep{chen2016open}, which contains 5,936 1mm thickness normal-dose 2D slices from 10 anonymous patients. Since only post-log projection data is available, we perform forward projection to generate the pre-log CT projection data based on fan-beam geometry and the corresponding parameters are presented in Table~\ref{tab:fan_beam}.
We choose 4,825 NDCT images from eight patients for the Gaussian diffusion model pretraining in DGDiff, along with the corresponding pre-log projections specifically for SPDiff training.
 1,111 images and pre-log projections from the remaining two patients are used as the test dataset.
To verify the generalization performance of the proposed method across unseen dose levels, we implement a more accurate simulation approach~\citep{zeng2015simple} by introducing ``Poisson + Gaussian'' noise into the pre-log raw measurements, generating data at 50\%, 25\%, 12.5\%, and 10\% dose levels. For this dataset, we set ${I_{0}}$ to $2.5\times 10^5$ and ${\sigma_{e}^{2}}$ to 10~\citep{xia2021ct}.

\noindent\textbf{Mayo 2020 dataset}\quad To further examine the generalization performance of our method on unseen dose levels from different datasets, we also choose the dataset \textit{Low Dose CT Image and Projection Data}~\citep{moen2021low} latest released by Mayo Clinic for testing. This dataset includes head and abdomen scans at 25\% dose and chest scans at 10\% dose.
Different from the Mayo 2016 dataset, we utilize the officially provided LDCT images and leverage the projection geometry outlined in Table~\ref{tab:fan_beam} to obtain the corresponding pre-log LDCT projection. As these LDCT images are simulated on post-log projections by adding signal-dependent Gaussian noise while incorporating bowtie filtration and automatic exposure control~\citep{yu2012development}, they provide valuable external validation that allows us to assess generalization performance across different noise distributions and scanning protocols.
We randomly select 5 chest scans and 5 abdomen scans, encompassing a total of 800 images that serve as the test set with mixed dose levels.

\begin{table}[!t]
\centering
\setlength{\abovecaptionskip}{3pt}
\caption{Parameters of fan-beam projection geometry.}
\label{tab:fan_beam}
\fontsize{8.5pt}{10pt}\selectfont
\begin{tabular*}{1\linewidth}{@{\extracolsep{\fill}}lc}
\shline
& Value \\
\midrule
Number of views &             672 \\
Number of detector bins &     672 \\
Distance between two contiguous rays (mm) &     2.72 \\
Distance between the source and rotation center (mm) &        746.02 \\
Distance between the detector and rotation center (mm) &      615.18 \\
\shline 
\end{tabular*}
\end{table}

\begin{figure*}[!t]
\centerline{
\includegraphics[width=1\textwidth]{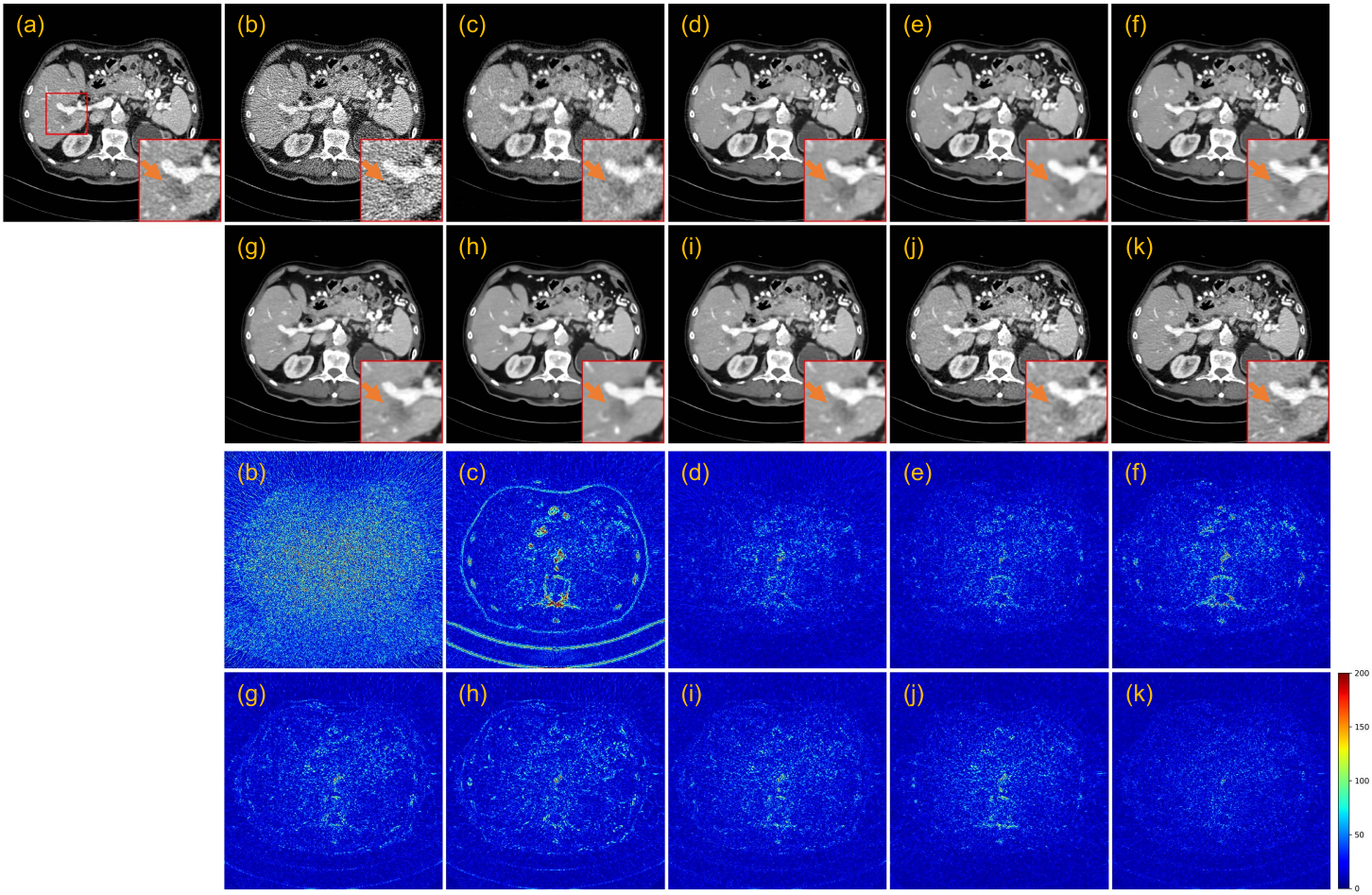}}
\caption{Qualitative results of a 25\% dose CT image from the Mayo 2016 dataset. (a) NDCT image (Ground truth), (b) FBP, (c) PWLS, (d) Noise2Noise, (e) Noise2Sim, (f) SSDDNet, (g) DR2, (h) GDP, (i) Dn-Dp, (j) \methodproj (\textbf{ours}), and (k) \methodname (\textbf{ours}). The display window is [-160, 240] HU. The red ROI is zoomed in for visual comparison and the orange arrow points to one lesion.}
\label{fig:mayo2016_25dose}
\end{figure*}

\subsection{Implementation Details}
The backbone of the shifted Poisson diffusion model is a U-Net with the input pre-log projection of size ${3\times672\times 672}$. Following~\citep{gao2022cocodiff}, we concatenate the pre-log projection slice and its adjacent slices in the channel dimension to leverage contextual information. The shifted Poisson diffusion model is trained using a batch size of 8 with 100k iterations. The learning rate is set to ${2\times 10^{-4}}$ and the total diffusion steps $\tau$ is set to 10. $\lambda_{t}$ is calculated as follows: 
$\lambda_{t}={\lambda_{0} \lambda_{\tau} \tau}/({\lambda_{\tau}(\tau-t)+\lambda_{0} t})$,
where we empirically set $\lambda_{1}$ to ${3\times 10^{5}}$ and $\lambda_{\tau}$ to ${2.5\times 10^{4}}$, to encompass a broad scenario involving the number of X-ray photons. The weighting parameter ${s}_{1}$ is set to $0.7 {{\exp(5000 {\sigma_\mathrm{ld}}^{2})}/({10+\exp(5000 {\sigma_\mathrm{ld}}^{2})})}$ and ${s}_{2}$ is set to $0.8$. The detailed derivation of the formula for $\lambda_{t}$ and the selection rule for ${s}_{1}$ and ${s}_{2}$ are given in Supp.~A.

Given the relatively high resolution of CT images at $512 \times 512$, training a Gaussian diffusion model for high-resolution image generation presents a significant challenge. Therefore, for this experiment, we employ a cascade framework in pretraining the Gaussian diffusion model in \methodimg, which has been proven effective in generating high-resolution images~\citep{ho2022cascaded, liu2023diffusion}. In the cascaded pipeline, we first train an unconditional diffusion model to generate CT images at a resolution of ${256\times 256}$, using a batch size of 16. Then, an additional conditional diffusion model is trained for ${256\times 256 \rightarrow 512\times 512}$ image super-resolution, with a smaller batch size of 4, employing the ${256\times 256}$ resolution CT images as the condition. The number of iterations in both stages is fixed at 700k, and the learning rate is set to ${8\times 10^{-5}}$. Total diffusion steps $T$ is ${1\times 10^{3}}$. During the sampling process of \methodimg, we utilize the proposed sampling pipeline in Eq.~\eqref{eq:image_domain_finetuning_pipeline} for $256\times 256$ image refinement. For ${256\times 256 \rightarrow 512\times 512}$ image super-resolution, we investigate two accelerated ODE samplers: DDIM~\citep{song2020denoising} and DPM-Solver~\citep{lu2022dpm}. Both samplers produced similar results, with DDIM showing slight advantages. We choose DDIM with 10 sampling steps and detailed results can be found in Supp.~B.
Both the shifted Poisson diffusion and Gaussian diffusion models are optimized by the Adam optimizer. We conduct all training and test experiments in the PyTorch platform with an NVIDIA RTX 4090 GPU (24GB). The training for SPDiff completed in approximately 11 hours, while for DGDiff, the Gaussian diffusion model for ${256\times 256}$ image generation required about 48 hours for training, and the conditional diffusion model for ${256\times 256 \rightarrow 512\times 512}$ image super-resolution needed 117 hours.
The simulation of LDCT data and forward projection is implemented based on the ASTRA toolbox~\citep{van2016fast}. In addition, we employ the TorchRadon toolbox~\citep{ronchetti2020torchradon} to speed up FBP reconstruction.

\subsection{Comparison Methods}
To evaluate the performance of the proposed \methodname, several paired ND/LDCT data-free methods are selected for comparison, including 1) iterative reconstruction algorithm: Penalized Weighted Least Squares model (PWLS)~\citep{wang2006penalized}; 2) self-supervised methods: Noise2Noise~\citep{lehtinen2018noise2noise}, Noise2Sim~\citep{niu2022noise}, and SSDDNet~\citep{niu2022self}; 3) unsupervised diffusion-based methods: DR2~\citep{wang2023dr2}, GDP~\citep{fei2023generative}, and Dn-Dp~\citep{liu2023diffusion}. Among them, Noise2Noise is not a self-supervised method in the strictest sense as it requires multiple paired noisy images for training. To this end, we simulate an additional set of 25\% and 10\% dose CT images from the Mayo 2016 dataset as training labels. Noise2Sim and SSDDNet are developed specifically for low-dose CT reconstruction only required LDCT data.
Noise2Sim is the image domain-based reconstruction model, while SSDDNet represents its dual-domain version.
DR2, GDP, and Dn-Dp are three image domain-based restoration algorithms using diffusion prior: DR2 leverages low-frequency information from degraded images as guidance for blind face restoration while GDP utilizes diffusion priors for diverse image restoration tasks. 
Since these diffusion-based methods rely on a pretrained Gaussian diffusion model as a prior and differ only in their sampling algorithms, we ensure a fair comparison by using the same Gaussian diffusion model pretrained on NDCT data consistent with our \methodname for all methods, with only the sampling process adjusted for each. The detailed inference procedure of comparative diffusion-based methods is given in Supp.~C
We implement all comparison methods referencing their original paper or official repositories.

\begin{figure*}[!t]
\centerline{
\includegraphics[width=1\textwidth]{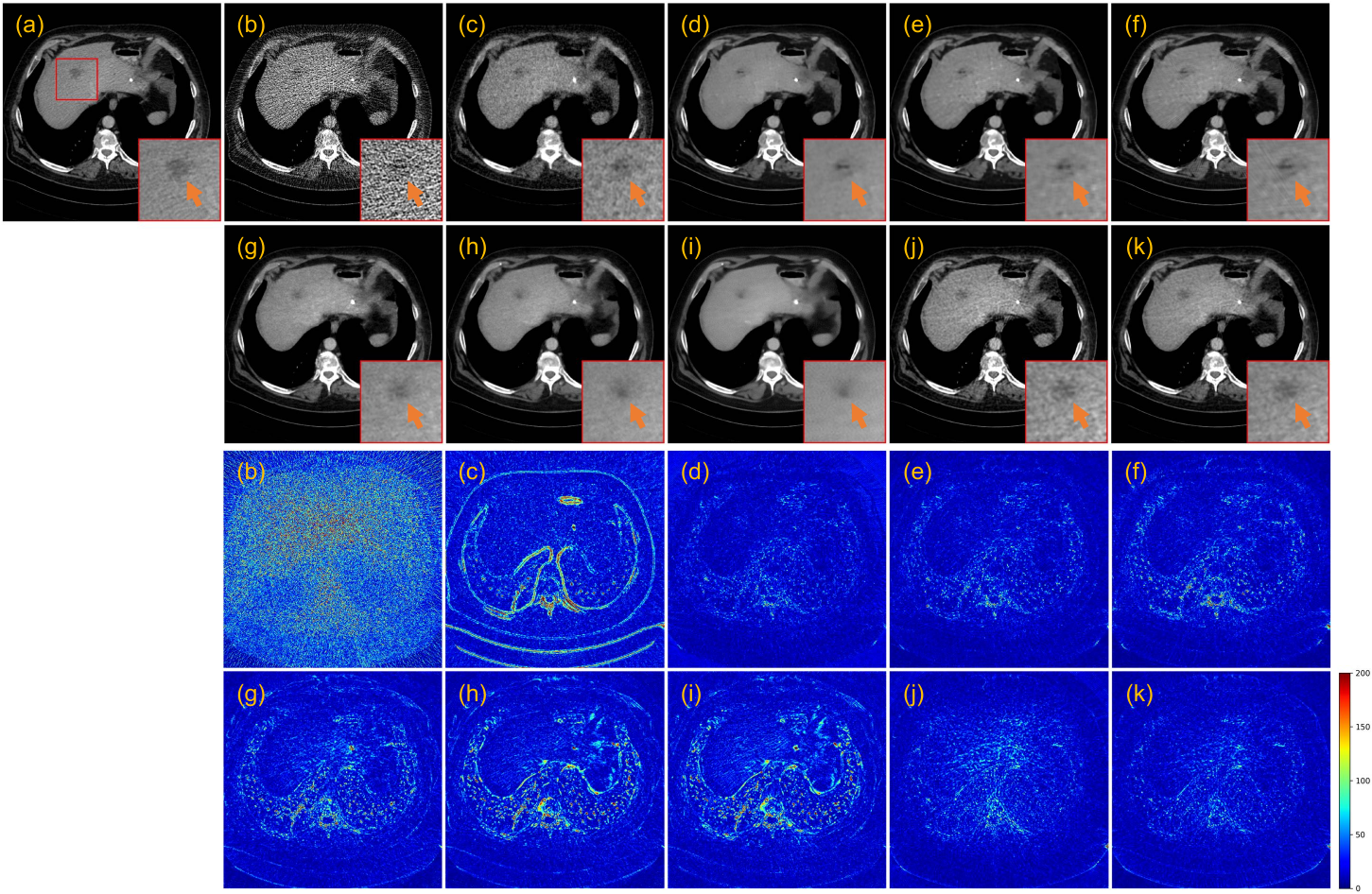}}
\caption{Qualitative results of a 10\% dose CT image from the Mayo 2016 dataset. (a) NDCT image (Ground truth), (b) FBP, (c) PWLS, (d) Noise2Noise, (e) Noise2Sim, (f) SSDDNet, (g) DR2, (h) GDP, (i) Dn-Dp, (j) \methodproj (\textbf{ours}), and (k) \methodname (\textbf{ours}). The display window is [-160, 240] HU. The red ROI is zoomed in for visual comparison and the orange arrow points to one lesion.}
\label{fig:mayo2016_10dose}
\end{figure*}

\subsection{Evaluation Metrics}
We use three popular image quality assessment metrics to evaluate the quantitative performance of these methods: peak signal-to-noise ratio (PSNR), structural similarity (SSIM) index~\citep{wang2004image}, and root mean square error (RMSE). For a more comprehensive evaluation, three metrics more closely related to radiologists' subjective perception are also included in the quantitative assessment~\citep{mason2019comparison}: feature similarity index (FSIM)~\citep{zhang2011fsim}, visual information fidelity (VIF)~\citep{sheikh2006image}, and noise quality metric (NQM)~\citep{damera2000image}. Better performance is indicated by higher PSNR, SSIM, FSIM, VIF, and NQM, or lower RMSE. We calculate all metrics using a CT window of [-1000, 1000] HU. To further evaluate the performance of these models for downstream tasks, we feed the reconstructed images into the foundation segmentation model MedSAM~\citep{ma2024segment} lesions and small tissues (\eg, bronchi and blood vessels) segmentation, whose gold standard is obtained from the segmentation results of NDCT images by MedSAM. Performance is quantified by calculating the Dice coefficient, intersection over union (IoU), and accuracy (Acc) between the segmentation results of the reconstructed images and the gold standard.

\subsection{Evaluation on Seen Doses and Dataset}
We evaluate the reconstruction performance of our method through both quantitative metrics and qualitative analysis on the 25\% and 10\% dose test data from the Mayo 2016 dataset. For DL-based methods, including the self-supervised and diffusion-based methods, the training dose matches the test dose for the former, while the latter are trained exclusively on NDCT data from the Mayo 2016 dataset.

\noindent\textbf{Evaluation on the 25\% dose}\quad Fig.~\ref{fig:mayo2016_25dose} presents the reconstruction results and residual maps of different methods on a representative slice with 25\% dose. All methods effectively reduce noise to varying extents; however, the traditional PWLS method induces a noticeable CT value drift at the tissue edges. Among the DL-based methods, Noise2Noise and SSDDNet, deliver superior denoising results but introduce streak artifacts. On the other hand, Noise2Sim and GDP, while effective, are prone to smooth tissue details. DR2 and Dn-Dp yield visually pleasing denoising outcomes. However, the examination of residual maps reveals obvious CT value drift due to the imprecise location of prior information, leading diffusion models to generate pixels deviating from the ground truth.
In contrast, our \methodproj reduces CT value bias by improving the diffusion process to align with the noise model in pre-log projection. Furthermore, benefiting from precise localization of prior information in the image domain, \methodname excels in both noise suppression and detail preservation, offering the optimal balance between the two. The red region of interest (ROI) also shows that our \methodname preserves the lesion well.

\begin{table*}[!t]
\centering
\caption{Quantitative results (mean$\pm$std) on seen dose levels from the Mayo 2016 dataset. The best results are highlighted in  \textbf{bold}.}
\label{tab:quantitative_results_on_25dose_dataset}
\setlength{\tabcolsep}{3pt}
\fontsize{6.9pt}{10pt}\selectfont
\begin{tabular*}{1\linewidth}{@{\extracolsep{\fill}}l|r|cccccc|cccccc}
\shline
\multirow{2}{*}{\textbf{Methods}} & Param & 
\multicolumn{6}{c|}{\textbf{25\% Dose}} & \multicolumn{6}{c}{\textbf{10\% Dose}}\\
& M)\hspace{1.0mm} & PSNR $\uparrow$ & SSIM(\%) $\uparrow$ & RMSE $\downarrow$ 
& FSIM(\%) $\uparrow$ & VIF(\%) $\uparrow$ & NQM $\uparrow$ 
& PSNR $\uparrow$ & SSIM(\%) $\uparrow$ & RMSE $\downarrow$ 
& FSIM(\%) $\uparrow$ & VIF(\%) $\uparrow$ & NQM $\uparrow$\\
\hline

FBP & - 
& 33.59$\pm$2.69 & 76.03$\pm$10.5 & 43.78$\pm$13.8 
& 95.30$\pm$2.41 & 59.70$\pm$8.99 & 32.32$\pm$3.68 
& 29.40$\pm$2.80 & 58.84$\pm$13.4 & 71.17$\pm$23.4 
& 90.29$\pm$4.51 & 46.65$\pm$8.72 & 28.16$\pm$3.86\\

PWLS & - 
& 36.28$\pm$0.66 & 92.42$\pm$1.86 & 30.67$\pm$2.17 
& 97.09$\pm$0.45 & 50.25$\pm$4.90 & 29.64$\pm$2.12
& 34.68$\pm$0.77 & 88.55$\pm$3.13 & 36.92$\pm$3.06 
& 95.80$\pm$0.67 & 43.35$\pm$5.20 & 26.35$\pm$2.45\\
\hline

Noise2Noise & 4.33
& 42.00$\pm$1.66 & 96.02$\pm$1.68 & 16.12$\pm$3.03
& 98.67$\pm$0.51 & 64.16$\pm$7.46 & 35.28$\pm$3.22
& 39.93$\pm$1.85 & 92.80$\pm$3.59 & 20.57$\pm$4.58 
& 97.97$\pm$0.73 & 56.70$\pm$7.12 & 31.93$\pm$3.35\\

Noise2Sim & 4.19
& 41.65$\pm$1.25 & 96.12$\pm$1.23 & 16.65$\pm$2.26 
& 98.52$\pm$0.38 & 60.72$\pm$5.83 & 35.10$\pm$2.50
& 40.08$\pm$1.32 & 95.15$\pm$1.57 & 19.97$\pm$2.87 
& 98.09$\pm$0.44 & 55.16$\pm$6.11 & 32.37$\pm$3.02\\

SSDDNet & 8.38
& 39.83$\pm$0.98 & 95.39$\pm$1.29 & 20.45$\pm$2.13 
& 98.17$\pm$0.34 & 56.12$\pm$4.74 & 33.82$\pm$2.06
& 38.42$\pm$1.22 & 93.61$\pm$2.11 & 24.16$\pm$3.34 
& 97.78$\pm$0.49 & 52.14$\pm$5.39 & 31.88$\pm$2.73\\
\hline

DR2 & 71.40
& 39.52$\pm$1.11 & 94.43$\pm$1.36 & 21.24$\pm$2.60 
& 97.96$\pm$0.35 & 52.57$\pm$4.12 & 32.21$\pm$1.88
& 38.34$\pm$1.35 & 94.04$\pm$1.54 & 24.42$\pm$4.18 
& 97.82$\pm$0.48 & 51.43$\pm$4.24 & 29.53$\pm$2.73\\

GDP & 71.40 
& 39.03$\pm$1.49 & 94.58$\pm$1.39 & 22.62$\pm$3.80 
& 97.58$\pm$0.56 & 52.03$\pm$4.74 & 29.84$\pm$2.09
& 38.35$\pm$1.51 & 94.31$\pm$1.57 & 24.48$\pm$4.37 
& 97.55$\pm$0.62 & 51.31$\pm$4.77 & 28.73$\pm$2.29\\

Dn-Dp & 71.40
& 40.62$\pm$1.09 & 95.48$\pm$1.18 & 18.70$\pm$2.21 
& 98.32$\pm$0.34 & 56.95$\pm$4.72 & 32.98$\pm$1.95
& 38.00$\pm$1.33 & 93.87$\pm$1.23 & 25.38$\pm$3.83 
& 97.52$\pm$0.61 & 50.36$\pm$4.46 & 29.19$\pm$2.15\\

\methodproj(\textbf{ours}) & 4.36  
& 41.75$\pm$1.57 & 96.40$\pm$1.33 & 16.55$\pm$2.88 
& 98.84$\pm$0.37 & 65.25$\pm$6.79 & 34.14$\pm$2.90
& 39.45$\pm$1.93 & 95.24$\pm$1.67 & 21.83$\pm$6.39 
& 98.21$\pm$0.66 & 61.81$\pm$6.38 & 30.34$\pm$3.27\\
\methodname (\textbf{ours}) & 75.76 
& \textbf{42.82$\pm$1.05} & \textbf{96.75$\pm$0.79} & \textbf{14.51$\pm$1.69} 
& \textbf{99.00$\pm$0.17} & \textbf{68.01$\pm$3.35} & \textbf{35.89$\pm$2.10}
& \textbf{40.90$\pm$1.67} & \textbf{96.07$\pm$1.26} & \textbf{18.34$\pm$4.61} 
& \textbf{98.73$\pm$0.49} & \textbf{64.66$\pm$5.10} & \textbf{31.65$\pm$3.12}\\
\shline 
\end{tabular*}
\end{table*}

\noindent\textbf{Evaluation on the 10\% dose}\quad Fig.~\ref{fig:mayo2016_10dose} presents the qualitative results of a representative slice with 10\% dose. 
PWLS, SSDDNet, DR2, GDP, and Dn-Dp all result in significant CT value drift due to increased noise and artifacts in the projections and FBP-reconstructed images. In contrast, Noise2Noise, Noise2Sim, and our \methodname better preserve CT values during reconstruction. However, when examining the lesion tissue within the zoomed-in red ROI, we find that Noise2Noise and Noise2Sim distort the lesion's structure, whereas our \methodname maintains the details and size of the lesion.

\begin{figure}[!t]
\centerline{
\includegraphics[width=0.5\textwidth]{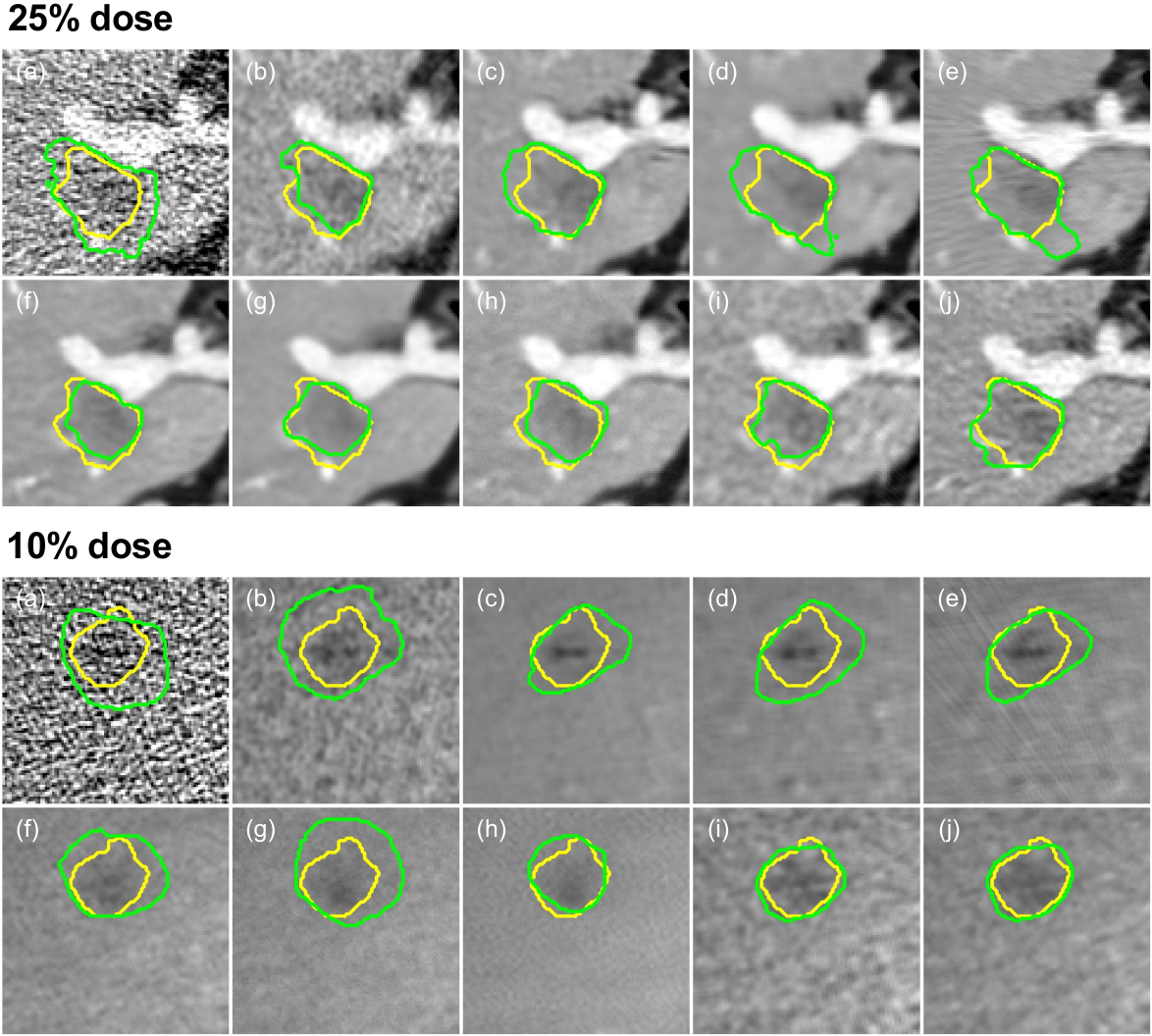}}
\caption{Visualization of lesion ROIs segmentation in Figs.~\ref{fig:mayo2016_25dose} and~\ref{fig:mayo2016_10dose}. Segmentation results are denoted in yellow for NDCT images and in green for images reconstructed by different methods. (a) FBP, (b) PWLS, (c) Noise2Noise, (d) Noise2Sim, (e) SSDDNet, (f) DR2, (g) GDP, (h) Dn-Dp, (i) \methodproj (\textbf{ours}), and (j) \methodname (\textbf{ours}). The display window is [-160, 240] HU.}
\label{fig:mayo2016_seg}
\end{figure}

\begin{table}[!t]
\centering
\caption{Quantitative evaluation of segmentation results on the Mayo 2016 dataset. The best results are highlighted in \textbf{bold}.}
\label{tab:segmentation_results_on_mayo2016_dataset}
\setlength{\tabcolsep}{3pt}
\fontsize{8pt}{10pt}\selectfont
\begin{tabular*}{1\linewidth}{@{\extracolsep{\fill}}l|ccc|ccc}
\shline
\multirow{2}{*}{\textbf{Methods}}    & \multicolumn{3}{c|}{\textbf{25\%Dose}}  & \multicolumn{3}{c}{\textbf{10\%Dose}}\\
& Dice $\uparrow$ & IoU $\uparrow$ & Acc(\%) $\uparrow$ 
& Dice $\uparrow$ & IoU $\uparrow$ & Acc(\%) $\uparrow$\\
\hline

FBP 
& 0.726 & 0.570 & 99.7490 
& 0.613 & 0.442 & 99.6777\\
PWLS 
& 0.829 & 0.708 & 99.8882
& 0.577 & 0.405 & 99.6052\\
\hline
Noise2Noise
& 0.846 & 0.733 & 99.8821 
& 0.778 & 0.637 & 99.8516\\
Noise2Sim 
& 0.822 & 0.697 & 99.8577 
& 0.687 & 0.524 & 99.7616\\
SSDDNet 
& 0.842 & 0.727 & 99.8775
& 0.754 & 0.605 & 99.8295\\
\hline
DR2 
& 0.842 & 0.728 & 99.9054 
& 0.721 & 0.564 & 99.7917\\
GDP 
& 0.861 & 0.756 & 99.9138 
& 0.616 & 0.445 & 99.6666\\
Dn-Dp 
& 0.844 & 0.731 & 99.9001 
& 0.843 & 0.729 & 99.9104\\
\methodproj (\textbf{ours}) 
& 0.857 & 0.750 & 99.9123 
& 0.861 & 0.757 & 99.9168\\
\methodname (\textbf{ours}) 
& \textbf{0.882} & \textbf{0.789} & \textbf{99.9149} 
& \textbf{0.863} & \textbf{0.760} & \textbf{99.9180}\\
\shline 
\end{tabular*}
\end{table}

\begin{figure*}[!t]
\centerline{
\includegraphics[width=1\linewidth]{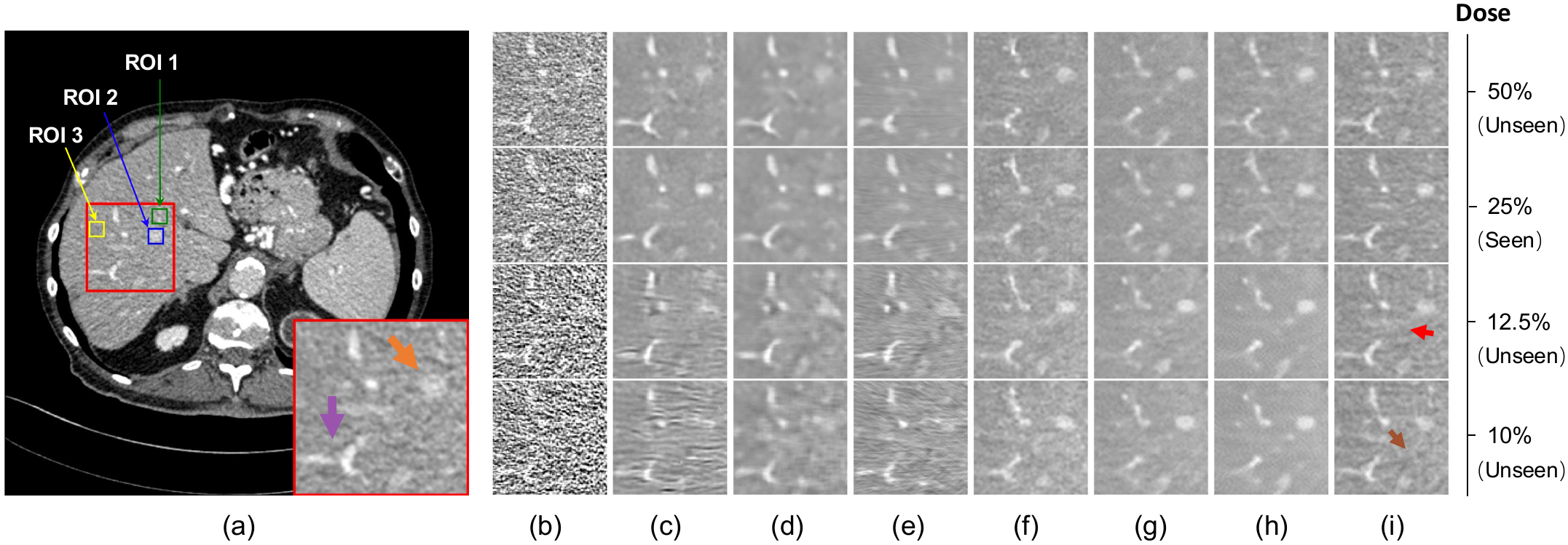}}
\caption{Qualitative results of a selected ROI across seen and unseen dose levels from the Mayo 2016 dataset. (a) NDCT image (Ground truth), (b) FBP, (c) Noise2Noise, (d) Noise2Sim, (e) SSDDNet, (f) DR2, (g) GDP, (h) Dn-Dp, and (i) \methodname (\textbf{ours}). The orange arrow and the purple arrow point to two key details for visual comparison across different methods, while the red and brown arrows indicate two key details for visual comparison of our \methodname across four dose levels. The green, blue, and yellow ROIs are selected for quantitative analysis.}
\label{fig:mayo2016_various_doses}
\end{figure*}

\noindent\textbf{Quantitative evaluation}\quad Table~\ref{tab:quantitative_results_on_25dose_dataset} presents the mean quantitative results and standard deviation across the entire test dataset for different methods, demonstrating that our \methodname outperforms all other self-supervised, unsupervised diffusion-based methods, and traditional iterative reconstruction methods, with \methodproj ranking second in overall performance. Table~\ref{tab:quantitative_results_on_25dose_dataset} also provides the number of parameters for all learning-based methods. The self-supervised methods have relatively fewer parameters. Among them, SSDDNet requires twice as many parameters as Noise2Sim because it processes LDCT data separately in both the projection and image domains. As for diffusion-based methods, DR2, GDP, and Dn-Dp maintain consistent parameter counts since they share the same Gaussian diffusion model as \methodimg. 
Diffusion-based models typically have more parameters than self-supervised methods, since their networks are designed for the more challenging NDCT image generation rather than specifically for LDCT reconstruction. However, their powerful modeling capabilities enable effective pretraining on NDCT data and allow for training-free use in various tasks such as data augmentation, image inpainting, and super-resolution. In this study, the Gaussian diffusion model is adapted for LDCT reconstruction and demonstrates excellent performance. Notably, our \methodproj reduces model size while delivering superior performance compared to other methods by aligning to the noise model of pre-log CT projection data. Finally, our \methodname achieves a parameter count comparable to other image domain diffusion models, even though it processes data in both domains.

\noindent\textbf{Evaluation on the downstream segmentation task}\quad 
Fig.~\ref{fig:mayo2016_seg} presents the segmentation result of lesion ROIs in Figs.~\ref{fig:mayo2016_25dose} and~\ref{fig:mayo2016_10dose} using the MedSAM. At the 25\% dose, all methods improve the segmentation performance of MedSAM compared to FBP reconstruction, with Noise2Sim and SSDDNet showing some over-segmentation. At the 10\% dose, segmentation accuracy obviously decreases for all methods except Dn-DP, \methodproj, and \methodname, due to distortion of the lesion structure in the reconstructed images, which interferes with MedSAM's inference. Despite increased noise and artifacts, segmentation results of MedSAM on \methodproj and \methodname reconstructions closely match those of NDCT, indicating effective preservation of lesion boundaries and CT values.
Table~\ref{tab:segmentation_results_on_mayo2016_dataset} further demonstrates that \methodname reconstructions achieve the best lesion preservation at both doses, with \methodproj following closely behind it. We also randomly select 50 slices from the Mayo 2016 dataset and extract 2 ROIs from each slice, resulting in a total of 100 ROIs for extended evaluation. This comprehensive testing demonstrates the robustness of our method in the downstream segmentation task across a wider range of scenarios, and the quantitative results can be found in Supp.~D.

\noindent\textbf{Inference time}\quad Table~\ref{tab:inference_time} compares the inference time of different diffusion-based methods. 
Since the ${256\times 256 \rightarrow 512\times 512}$ stage focuses on image super-resolution and all methods employ the DDIM sampler (with 29 sampling steps for Dn-DP and 10 steps for other methods), our analysis lies on their computational efficiency when sampling on ${256\times 256}$ resolution images, which is the primary stage for refinement.
Despite sampling across dual domains, our method reduces inference time by approximately 46\% compared to DR2, 75\% compared to GDP, and 66\% compared to Dn-Dp for the 25\% dose case. 
Furthermore, even in the 10\% dose scenario, our method maintains a lower overall time consumption than all comparison methods.

\begin{table}[!t]
\centering
\caption{Inference times of diffusion-based methods.}
\label{tab:inference_time}
\fontsize{9pt}{10pt}\selectfont
\setlength{\tabcolsep}{3pt}
\begin{tabular*}{1\linewidth}{@{\extracolsep{\fill}}lccc}
\shline
& Projection & Image ($256\times256$) & Overall\\
\hline
DR2 &                     - & 0.377s & 0.377s\\
GDP &                     - & 0.813s & 0.813s\\
Dn-Dp &                   - & 0.610s & 0.610s\\
\methodname (25\%) & 0.084s & 0.121s & 0.205s\\
\methodname (10\%) & 0.108s & 0.142s & 0.250s\\
\shline 
\end{tabular*}
\end{table}

\begin{figure*}[!t]
\centerline{
\includegraphics[width=1\textwidth]{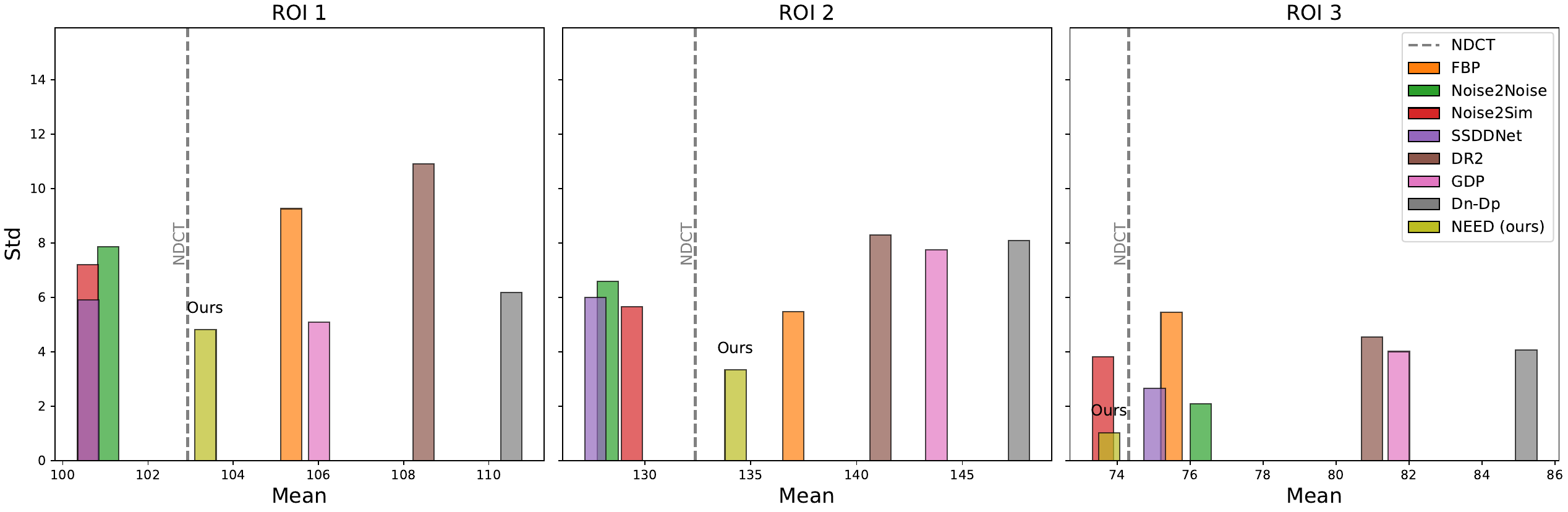}}
\caption{Mean CT values and standard deviation (std) of each ROI in Fig.~\ref{fig:mayo2016_various_doses} across four dose levels. The vertical axis represents the standard deviation (lower is better), and the horizontal axis represents the mean (closer to the NDCT, indicated by the dashed line, is better).}
\label{fig:mean_sd}
\end{figure*}

\begin{figure*}[!t]
\centerline{\includegraphics[width=1\linewidth]{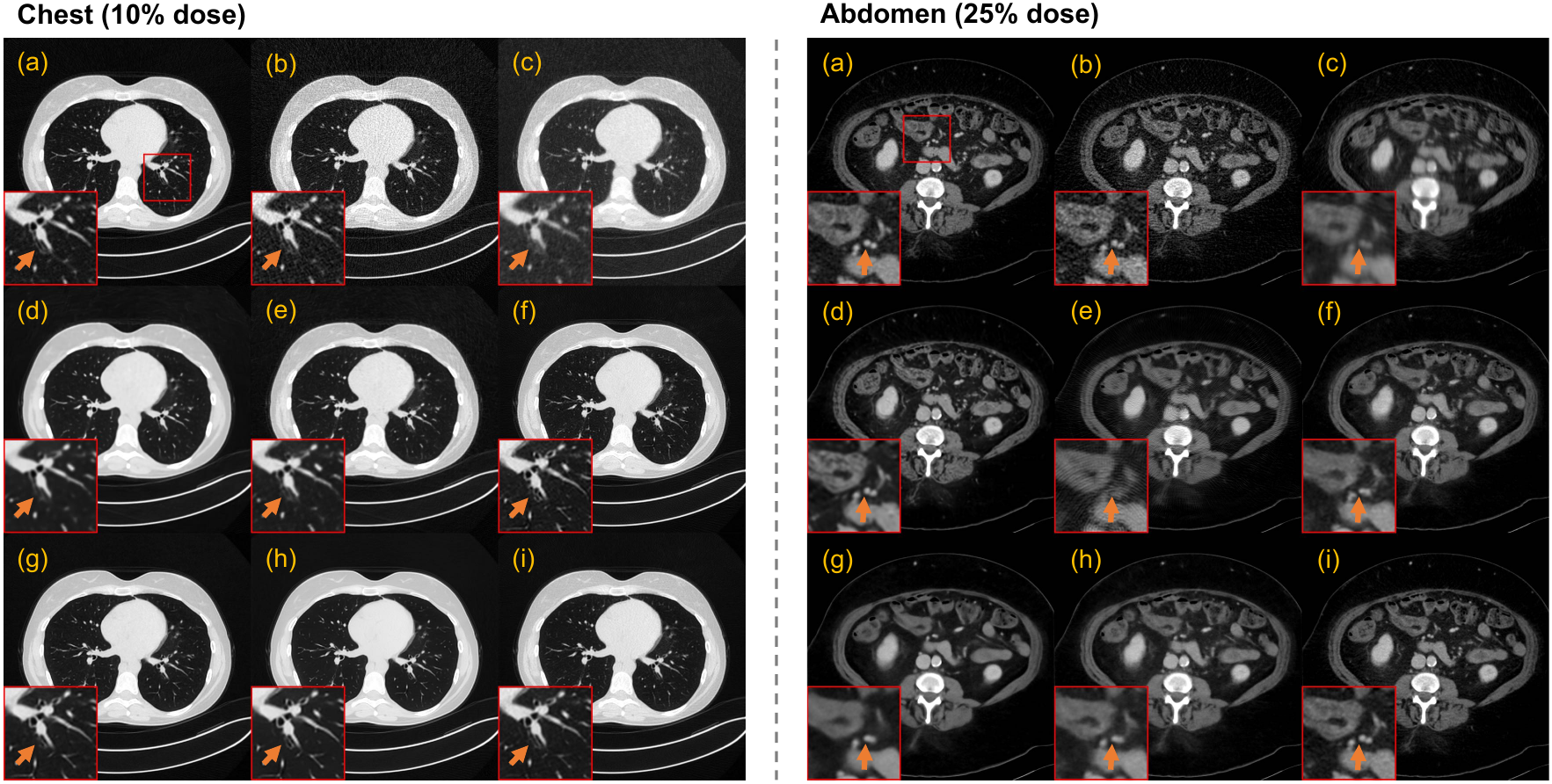}}
\caption{Qualitative results on the unseen Mayo 2020 dataset. (a) NDCT image (Ground truth), (b) FBP, (c) PWLS, (d) Noise2Sim, (e) SSDDNet, (f) DR2, (g) GDP, (h) Dn-Dp, and (i) \methodname (\textbf{ours}). The display window for chest images is [-1350, 150] HU and for abdomen images is [-160, 240] HU. The red ROI is zoomed in for visual comparison and the orange arrow points to a key detail.}
\label{fig:mayo2020}
\end{figure*}

\begin{table*}[t]
\centering
\caption{Quantitative results (mean$\pm$std) on Mayo 2020 dataset. The best results are highlighted in  \textbf{bold}.}
\label{tab:quantitative_results_on_mayo2020_dataset}
\setlength{\tabcolsep}{3pt}
\fontsize{8pt}{10pt}\selectfont
\begin{tabular*}{1\linewidth}{@{\extracolsep{\fill}}l|ccc|ccc|ccc}
\shline
\multirow{2}{*}{\textbf{Methods}}  & \multicolumn{3}{c|}{\textbf{Chest (10\%dose)}}  & \multicolumn{3}{c|}{\textbf{Abdomen (25\%dose)}}  & \multicolumn{3}{c}{\textbf{All (mixed dose levels)}}\\
& PSNR $\uparrow$ & SSIM(\%) $\uparrow$ & RMSE $\downarrow$ 
& PSNR $\uparrow$ & SSIM(\%) $\uparrow$ & RMSE $\downarrow$ 
& PSNR $\uparrow$ & SSIM(\%) $\uparrow$ & RMSE $\downarrow$ \\
\hline

FBP 
& 27.87$\pm$1.54 & 56.18$\pm$5.70 & 82.05$\pm$14.5
& 39.88$\pm$3.46 & 95.89$\pm$3.96 & 21.57$\pm$9.65
& 33.87$\pm$6.58 & 76.03$\pm$20.5 & 51.81$\pm$32.7\\

PWLS 
& 29.92$\pm$1.19 & 72.07$\pm$2.78 & 64.38$\pm$8.95
& 39.51$\pm$3.41 & 97.30$\pm$1.44 & 22.38$\pm$9.32
& 34.72$\pm$5.44 & 84.68$\pm$12.8 & 43.38$\pm$22.9\\
\hline

Noise2Noise 
& 31.90$\pm$1.47 & 76.53$\pm$3.85 & 51.49$\pm$8.61
& 40.88$\pm$3.49 & 98.05$\pm$1.11 & 19.19$\pm$8.11
& 36.39$\pm$5.23 & 87.29$\pm$11.1 & 35.34$\pm$18.2\\

Noise2Sim 
& 32.39$\pm$1.20 & 78.91$\pm$2.74 & 48.46$\pm$6.58
& 40.62$\pm$3.32 & 97.96$\pm$1.05 & 19.63$\pm$7.92
& 36.50$\pm$4.81 & 88.44$\pm$9.76 & 34.04$\pm$16.2\\

SSDDNet 
& 32.18$\pm$1.13 & 78.93$\pm$2.93 & 49.60$\pm$6.35
& 38.89$\pm$2.83 & 96.01$\pm$2.75 & 23.48$\pm$8.00
& 35.53$\pm$3.99 & 87.47$\pm$9.01 & 36.54$\pm$14.9\\
\hline

DR2 
& 31.46$\pm$0.90 & 76.06$\pm$3.50 & 53.70$\pm$5.48
& 38.48$\pm$2.12 & 96.37$\pm$1.30 & 24.09$\pm$6.35
& 34.97$\pm$3.87 & 86.22$\pm$10.5 & 38.89$\pm$16.0\\

GDP 
& 32.55$\pm$1.07 & 81.58$\pm$2.85 & 47.49$\pm$5.68
& 38.42$\pm$2.02 & 96.57$\pm$1.08 & 24.13$\pm$5.52
& 35.49$\pm$3.35 & 89.08$\pm$7.81 & 35.81$\pm$13.0\\

Dn-Dp 
& 31.66$\pm$0.94 & 76.68$\pm$2.85 & 52.52$\pm$5.54
& 39.44$\pm$2.62 & 96.99$\pm$1.22 & 21.89$\pm$7.09
& 35.55$\pm$4.36 & 86.83$\pm$10.4 & 37.21$\pm$16.6\\

\methodproj(\textbf{ours}) 
& 32.59$\pm$1.24 & 78.75$\pm$3.06 & 47.36$\pm$6.59
& 40.96$\pm$3.29 & 98.08$\pm$0.99 & 18.81$\pm$7.38
& 36.78$\pm$4.87 & 88.42$\pm$9.93 & 33.09$\pm$15.9\\

\methodname(\textbf{ours}) 
& \textbf{33.51$\pm$1.24} & \textbf{82.38$\pm$3.38} & \textbf{42.61$\pm$6.22}
& \textbf{41.80$\pm$2.80} & \textbf{98.23$\pm$0.93} & \textbf{16.83$\pm$6.11}
& \textbf{37.65$\pm$4.67} & \textbf{90.30$\pm$8.30} & \textbf{29.72$\pm$14.3}\\
\shline 
\end{tabular*}
\end{table*}

\subsection{Evaluation on Unseen Doses and Dataset}
To effectively evaluate the generalizable reconstruction performance of our \methodname, we extend the experiments to \textit{1) unseen dose levels from the same dataset} (Mayo 2016 dataset) and \textit{2) unseen dose levels from the unseen dataset} (Mayo 2020 dataset). All DL-based models are trained using data from the Mayo 2016 dataset, while Noise2Noise, Noise2Sim, and SSDDNet are trained using only 25\% dose data. DR2, GDP, Dn-Dp, and our \methodname are trained solely on NDCT data. It should be noted that due to differences in acquisition protocols across datasets, the 25\% dose from the Mayo 2020 dataset also slightly differs from that of the Mayo 2016 dataset.

\noindent\textbf{Unseen dose levels from the same dataset}\quad Fig.~\ref{fig:mayo2016_various_doses} shows the qualitative results of a selected ROI by different methods on the four dose levels from the Mayo 2016 dataset, namely 50\%, 25\%, 12.5\%, and 10\%. At 25\% and higher 50\% doses, all methods demonstrate adequate noise suppression performance.
However, when the test dose falls below 25\%, the denoising efficacy of all models, except unsupervised diffusion-based methods, experiences a sharp decline. Specifically, Noise2Noise and SSDDNet introduce noticeable streak artifacts, whereas Noise2Sim tends to blur the low-contrast tissue pointed by the orange arrow. Unsupervised diffusion-based methods are less sensitive to variations in test dose levels, owing to their exclusive training on NDCT data. Among them, Dn-Dp produces some speckle artifacts. GDP blurs the blood vessel indicated by the purple arrow, whereas DR2 performs better and preserves it by using low-frequency information to guide the sampling process, focusing on high-frequency details restoration. Our \methodname delivers optimal results by comprehensively examining texture and tissue details across four dose levels. However, the reconstructed image at 12.5\% dose shows noticeable blurring of vessels (indicated by the red arrow), while the reconstructed image at 10\% dose exhibits artifacts (marked by the brown arrow) that are successfully suppressed in higher-dose reconstructions. These findings demonstrate that while our \methodname delivers consistent reconstructions across various dose levels, performance remains dose-dependent.

We select two ROIs containing blood vessels (1 and 2) and one background ROI (3) to calculate the mean CT value and standard deviation across four dose levels, as presented in Fig.~\ref{fig:mean_sd}. Our \methodname yields the mean CT value closest to NDCT and the lowest standard deviation across all three ROIs, which demonstrates that our \methodname not only preserves CT values effectively but also exhibits superior robustness when tested on unseen dose levels.
Interestingly, while diffusion-based comparative methods offer visual stability, they are hindered by using LDCT images as guidance, leading to inaccurate prior information location and the generation of unrealistic pixels. In contrast, our method achieves high-quality initial reconstruction due to better alignment with the noise model in the projection domain, while doubly guidance in the image domain enhances data fidelity.

\begin{figure}[!t]
\centerline{
\includegraphics[width=0.5\textwidth]{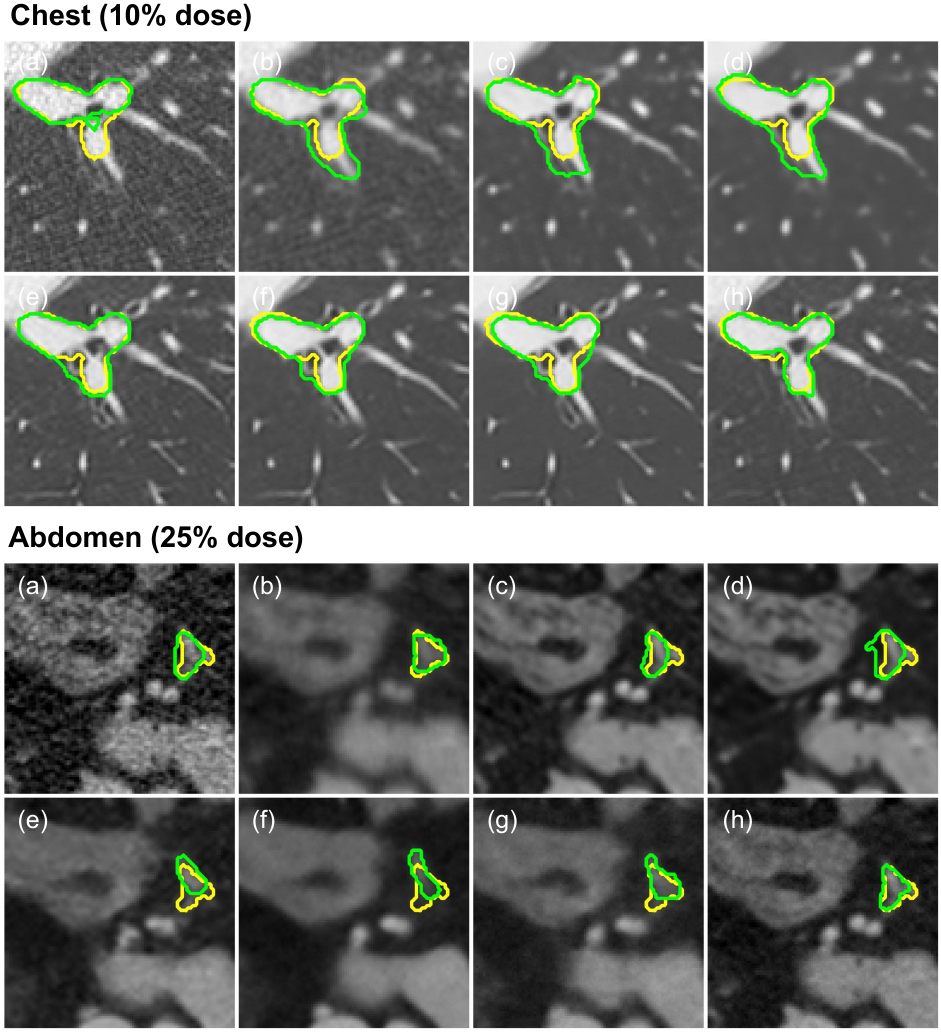}}
\caption{Visualization of ROIs segmentation in Fig.~\ref{fig:mayo2020}. Segmentation results are denoted in yellow for NDCT images and in green for images reconstructed by different methods. (a) FBP, (b) PWLS, (c) Noise2Noise, (d) Noise2Sim, (e) DR2, (f) GDP, (g) Dn-Dp, and (h) \methodname (\textbf{ours}). The display window for chest ROIs is [-1350, 150] HU and for abdomen ROIs is [-160, 240] HU.}
\label{fig:mayo2020_seg}
\end{figure}

\begin{table}[t]
\centering
\caption{Quantitative evaluation of segmentation results on the Mayo 2020 dataset. The best results are highlighted in \textbf{bold}.}
\label{tab:segmentation_results_on_mayo2020_dataset}
\setlength{\tabcolsep}{3pt}
\fontsize{8pt}{10pt}\selectfont
\begin{tabular*}{1\linewidth}{@{\extracolsep{\fill}}l|ccc|ccc}
\shline
\multirow{2}{*}{\textbf{Methods}}    & \multicolumn{3}{c|}{\textbf{Chest (10\%Dose)}}  & \multicolumn{3}{c}{\textbf{Abdomen (25\%Dose)}}\\
& Dice $\uparrow$ & IoU $\uparrow$ & Acc(\%) $\uparrow$ 
& Dice $\uparrow$ & IoU $\uparrow$ & Acc(\%) $\uparrow$\\
\hline
FBP 
& 0.846 & 0.733 & 99.9184 
& 0.786 & 0.647 & 99.9744\\
PWLS 
& 0.839 & 0.723 & 99.8928 
& 0.831 & 0.710 & 99.9798\\
\hline
Noise2Noise
& 0.838 & 0.721 & 99.8920
& 0.736 & 0.582 & 99.9710\\
Noise2Sim 
& 0.844 & 0.731 & 99.8955
& 0.669 & 0.502 & 99.9565\\
\hline
DR2 
& 0.915 & 0.843 & 99.9462
& 0.567 & 0.396 & 99.9557\\
GDP 
& 0.897 & 0.814 & 99.9382
& 0.587 & 0.416 & 99.9577\\
Dn-Dp 
& 0.874 & 0.776 & 99.9237
& 0.739 & 0.586 & 99.9706\\
\methodname (\textbf{ours}) 
& \textbf{0.924} & \textbf{0.859} & \textbf{99.9572} 
& \textbf{0.845} & \textbf{0.732} & \textbf{99.9832}\\
\shline 
\end{tabular*}
\end{table}

\begin{figure*}[!t]
\centerline{
\includegraphics[width=1\textwidth]{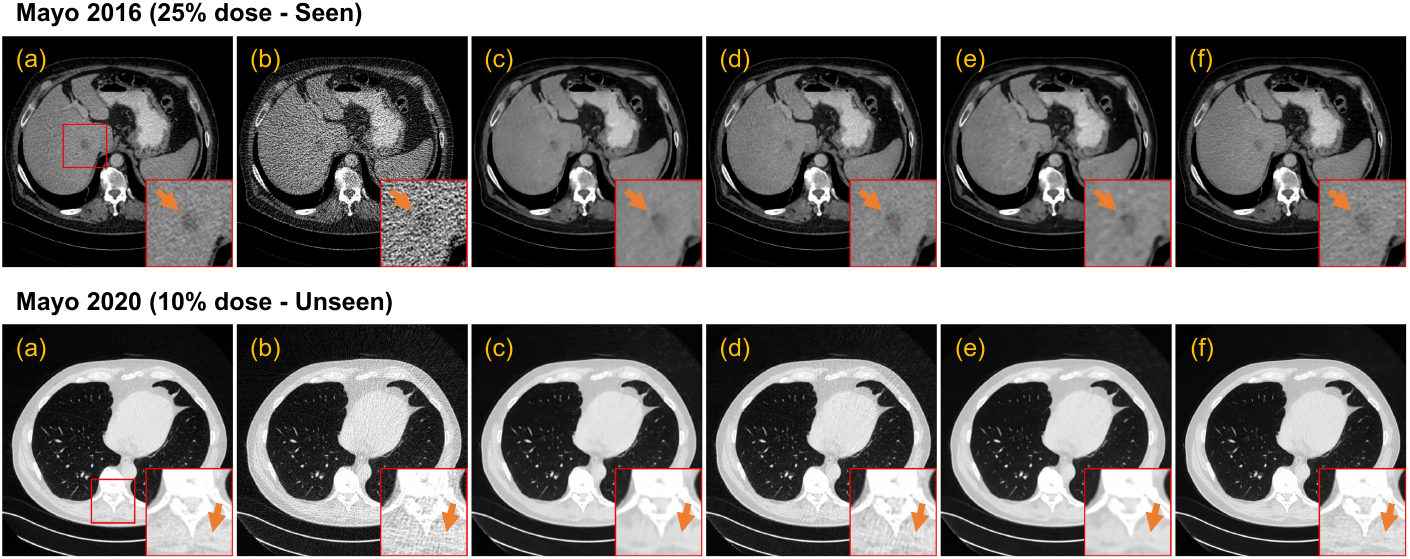}}
\caption{Qualitative comparison of supervised methods, Noise2Sim, and NEED on seen and unseen doses. (a) NDCT image (Ground truth), (b) FBP, (c) RED-CNN, (d) CoreDiff, (e) Noise2Sim, and (f) \methodname (\textbf{ours}). The display window for chest images is [-1350, 150] HU and for abdomen images is [-160, 240] HU. The red ROI is zoomed in for visual comparison and the orange arrow points to a key detail}.
\label{fig:supervised_img}
\end{figure*}

\begin{figure*}[!t]
\centerline{
\includegraphics[width=1\textwidth]{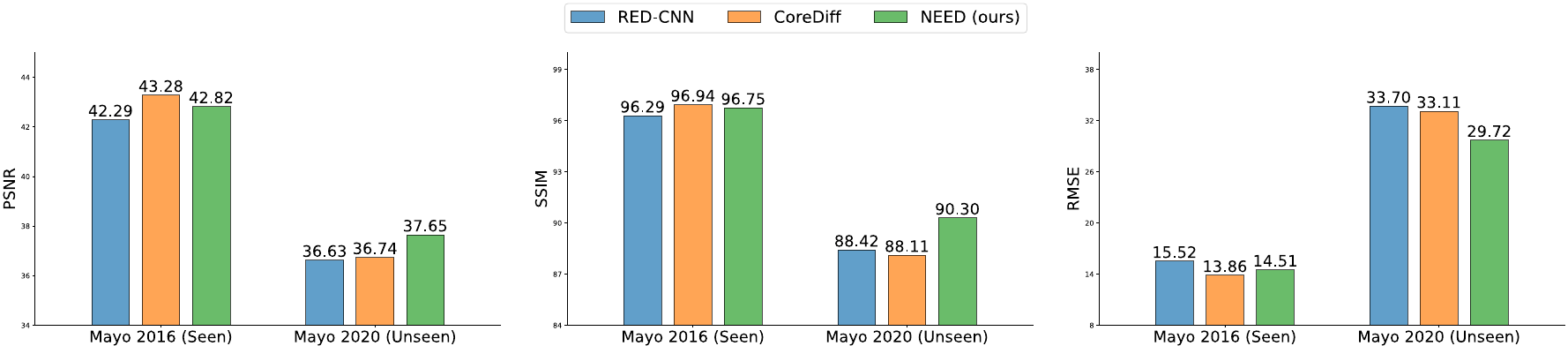}}
\caption{{Quantitative comparison of RED-CNN, CoreDiff, and \methodname (\textbf{ours}) on seen and unseen datasets.}}
\label{fig:supervised_hist}
\end{figure*}

\noindent\textbf{Unseen dose levels from the unseen dataset}\quad We further assess the effectiveness of our \methodname on the 25\% and 10\% mixed dose levels from the Mayo 2020 dataset.
Fig.~\ref{fig:mayo2020} presents the reconstruction results of different methods on two representative slices for visual comparison. The image reconstructed by PWLS exhibits velvet artifacts in the chest slice and blurs in the abdomen slice. Among the DL-based methods, Noise2Sim, SSDDNet, GDP, and DR2 tend to oversmooth the images, with SSDDNet also introducing ring artifacts in the abdomen slices. In contrast, DR2 and our \methodname yield sharper results. However, diffusion-based comparative methods including DR2 fail to preserve the key detail indicated by the orange arrow and introduce unrealistic structures.
Finally, our \methodname delivers visually optimal results.
Table~\ref{tab:quantitative_results_on_mayo2020_dataset} shows the quantitative evaluation of various methods. 
While \methodproj generalizes effectively to the Mayo 2020 dataset, its focus on denoising within the projection domain results in lower SSIM and higher RMSE metrics compared to GDP at 10\% dose, but it outperforms GDP at 25\% dose. After refinement in the image domain, \methodname outperforms all other methods in all terms. Fig.~\ref{fig:mayo2020_seg} and Table~\ref{tab:segmentation_results_on_mayo2020_dataset} also present both qualitative and quantitative evaluations of the segmentation results of ROIs using MedSAM, demonstrating that the reconstructed images of our NEED can effectively enhance the performance of downstream segmentation tasks, even when facing unseen dose levels and datasets. Similar to the experiments on the Mayo 2016 dataset, we further randomly select 100 ROIs from this dataset to demonstrate the robustness of our \methodname in the downstream segmentation task. The quantitative results are also provided in Supp.~D.

\subsection{Comparison with Supervised Methods}
In this section, we compare our method with two representative supervised approaches: the CNN-based RED-CNN~\citep{chen2017low} and the diffusion-based CoreDiff~\citep{gao2024corediff}. 
RED-CNN and CoreDiff are trained on paired 25\% and normal dose CT images from the Mayo 2016 dataset. Consistent with the previous experiments, we first evaluate denoising performance on the seen 25\% dose test images from this dataset, followed by generalization testing on the unseen dose levels from the unseen Mayo 2020 dataset.

Fig.~\ref{fig:supervised_img} presents the qualitative comparison of two supervised methods, the self-supervised method Noise2Sim, and our method.
When tested on the seen 25\% dose from the Mayo 2016 dataset, all methods effectively reduce noise, RED-CNN exhibits over-smoothing, whereas CoreDiff and \methodname retain more texture details. 
However, supervised methods degrade in performance when tested on an unseen 10\% dose from the unseen dataset. The over-smoothing in RED-CNN is further exacerbated, while CoreDiff struggles with inadequate noise suppression. In contrast, our \methodname effectively preserves bone edge details, offering more robust reconstruction performance.
Fig.~\ref{fig:supervised_hist} further validates our findings. Despite being trained exclusively on NDCT data, our method outperforms RED-CNN across all metrics at known dose levels. When tested on unseen dose levels from the Mayo 2020 dataset, \methodname exceeds both RED-CNN and CoreDiff.

\begin{figure*}[!t]
\centerline{\includegraphics[width=1\linewidth]{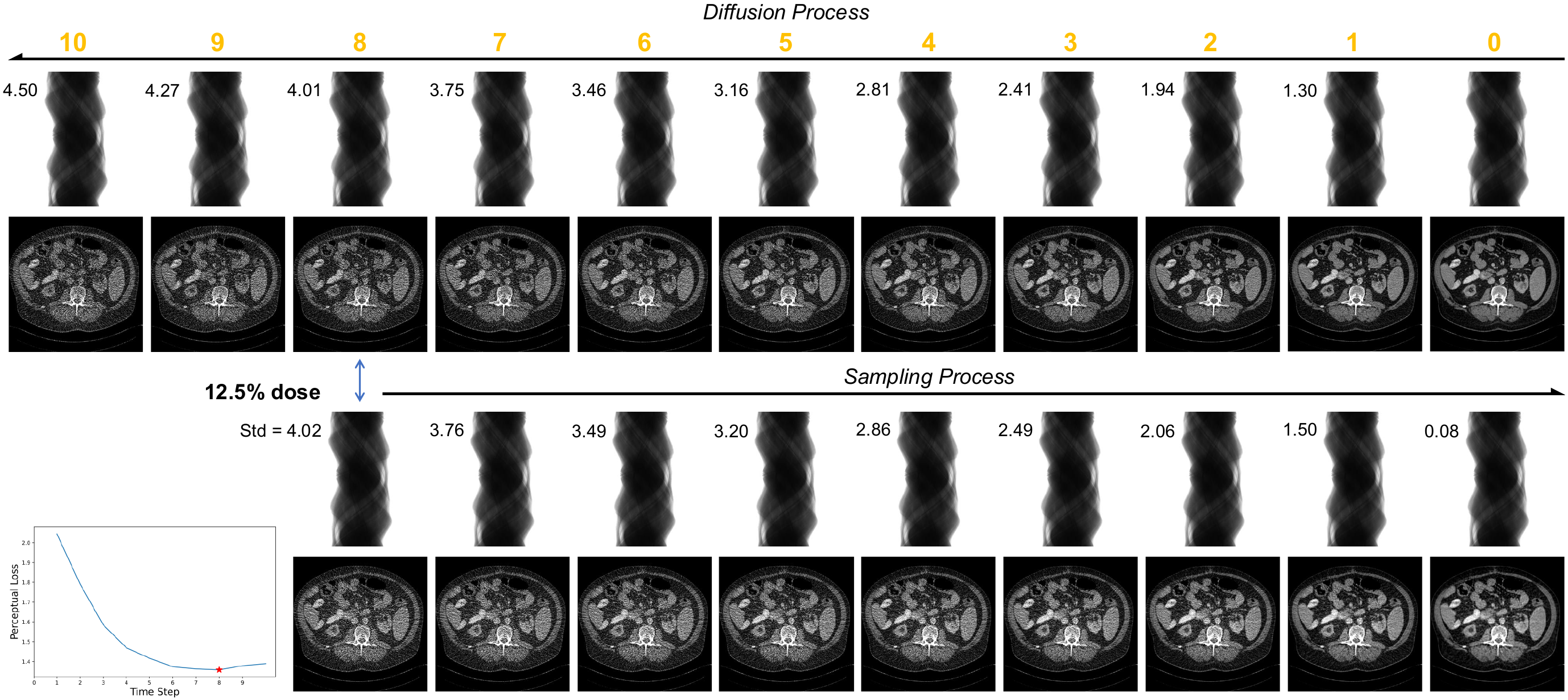}}
\caption{Visualization of diffusion and sampling processes along with time step matching strategy for the \methodproj. The dose of the example LDCT data is 12.5\%, with the perceptual loss between it and intermediate reconstructed images of the diffusion process shown in the lower left corner. We also provide the standard deviation ($\times 10^{-3}$) alongside each projection to indicate its noise level.}
\label{fig:timestep_matching}
\end{figure*}

\begin{figure*}[!t]
\centerline{
\includegraphics[width=1\linewidth]{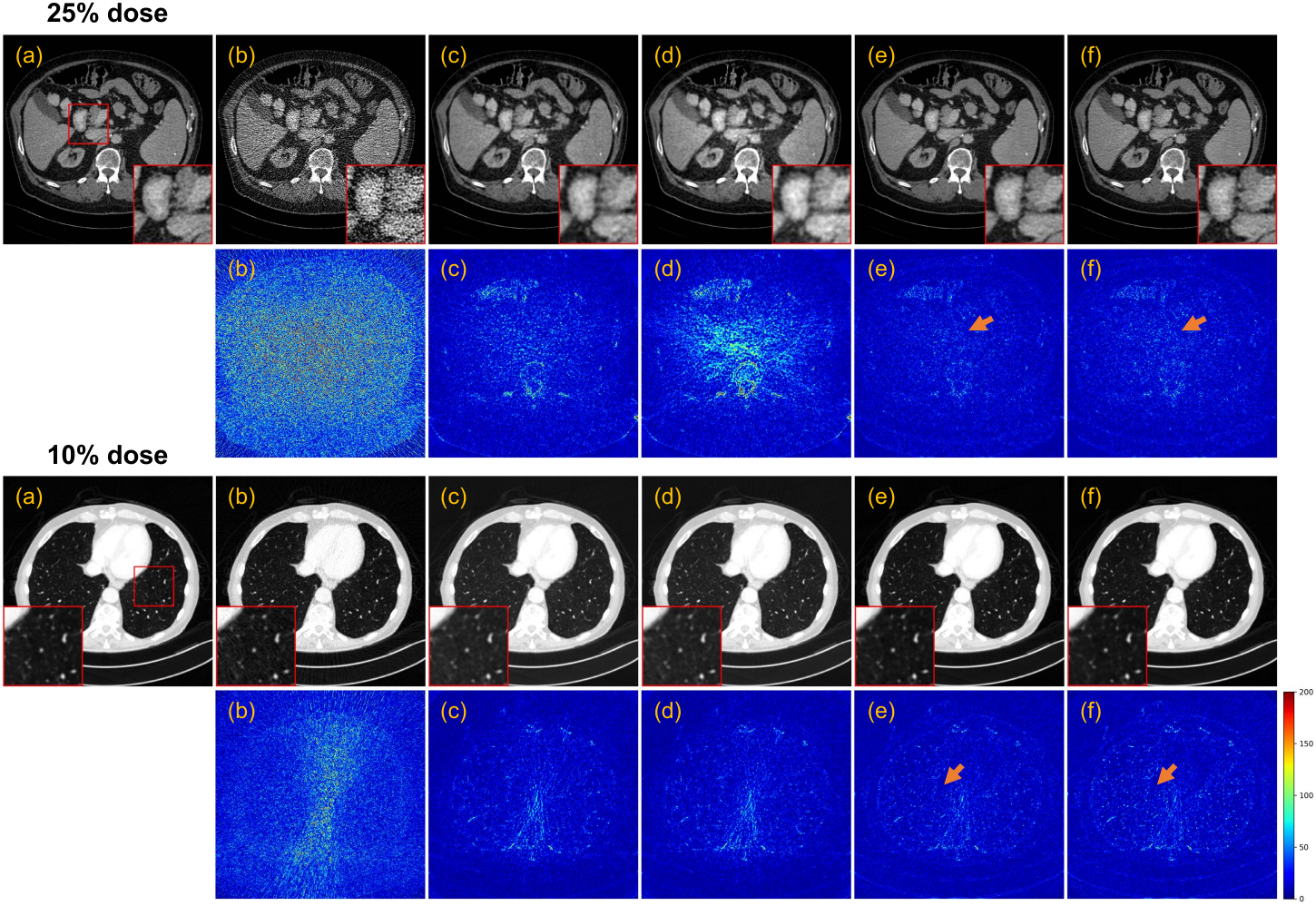}}
\caption{Qualitative results of our \methodproj and \methodname starting sampling from different time steps on the Mayo 2016 dataset. (a) NDCT image (Ground truth), (b) FBP, (c) \methodproj ($t_\mathrm{prj}^{*}$) (\textbf{ours}), (d) \methodproj ($\tau$), (e) \methodname ($t_\mathrm{prj}^{*}+t_\mathrm{img}^{*}$) (\textbf{ours}), and (f) \methodname ($t_\mathrm{prj}^{*}+T$). The display window for chest images is [-1350, 150] HU and for abdomen images is [-160, 240] HU. The red ROI is zoomed in for visual comparison and the orange arrow points to the key region in the residual map used for comparing \methodname ($t_\mathrm{prj}^{*}+t_\mathrm{img}^{*}$) and \methodname ($t_\mathrm{prj}^{*}+T$).}
\label{fig:mayo2016_all_t}
\end{figure*}

\begin{table*}[!t]
\centering
\caption{Quantitative results (mean$\pm$std) and inference time of \methodproj and \methodname starting sampling from different time steps on the Mayo 2016 dataset.
}
\label{tab:mayo2016_all_t}
\setlength{\tabcolsep}{3pt}
\fontsize{8.5pt}{10pt}\selectfont
\begin{tabular*}{1\linewidth}{@{\extracolsep{\fill}}l|cccc|cccc}
\shline
\multirow{2}{*}{\textbf{Methods}}
& \multicolumn{4}{c|}{\textbf{25\% dose}}  
& \multicolumn{4}{c}{\textbf{10\% dose}}\\
& PSNR $\uparrow$ & SSIM(\%) $\uparrow$ & RMSE $\downarrow$ & Time (s) 
& PSNR $\uparrow$ & SSIM(\%) $\uparrow$ & RMSE $\downarrow$ & Time (s)\\
\hline

\methodproj ($t_\mathrm{prj}^{*}$) (\textbf{ours})
& 41.75$\pm$1.57 & 96.40$\pm$1.33 & 16.55$\pm$2.88 & 0.084
& 39.45$\pm$1.93 & 95.24$\pm$1.67 & 21.83$\pm$6.39 & 0.108\\

\methodproj ($\tau$)
& 39.46$\pm$1.88 & 95.77$\pm$1.46 & 21.71$\pm$4.70 & 0.108
& 39.45$\pm$1.93 & 95.24$\pm$1.67 & 21.83$\pm$6.39 & 0.108\\
\hline
\methodname ($t_\mathrm{prj}^{*}+t_\mathrm{img}^{*}$) (\textbf{ours})
& 42.82$\pm$1.05 & 96.75$\pm$0.79 & 14.51$\pm$1.69 & 0.121
& 40.90$\pm$1.67 & 96.07$\pm$1.26 & 18.35$\pm$4.61 & 0.142\\

\methodname ($t_\mathrm{prj}^{*}+T$)
& 42.59$\pm$1.20 & 96.63$\pm$0.93 & 14.93$\pm$2.01 & 15.97
& 40.65$\pm$1.62 & 95.85$\pm$1.27 & 18.85$\pm$4.61 & 15.97\\
\shline 
\end{tabular*}
\end{table*}

\begin{figure}[!t]
\centerline{\includegraphics[width=1\columnwidth]{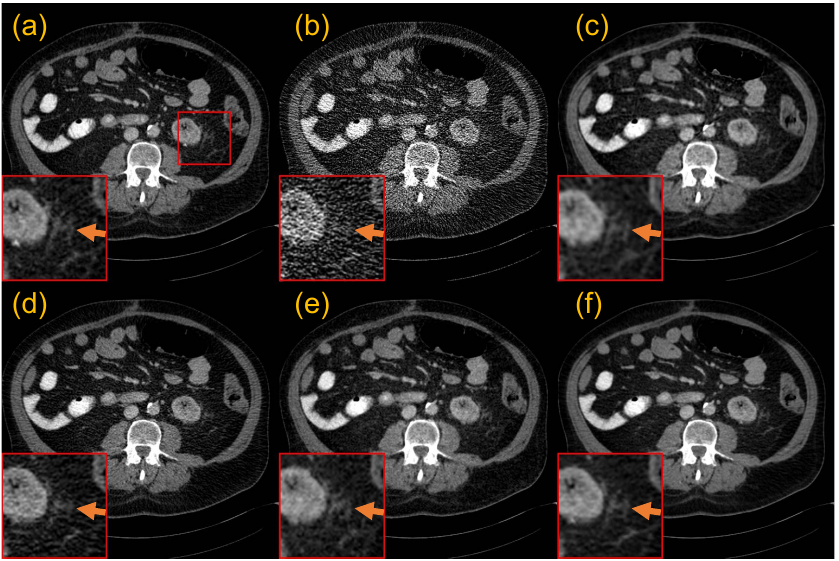}}
\caption{Qualitative analysis of proposed modules on 25\% dose test data from the Mayo 2016 dataset. (a) NDCT image (Ground truth), (b) FBP, (c) \methodproj, (d) SGDiff, (e) SPDiff+SGDiff, and (f) \methodname. The display window is [-160, 240] HU. }
\label{fig:ablation_25dose}
\end{figure}

\begin{table}[!t]
\centering
\caption{Ablation studies (mean$\pm$std) on the proposed modules on 25\% dose test data from the Mayo 2016 dataset.}
\label{tab:quantitative_results_on_ablation_25dose}
\setlength{\tabcolsep}{3pt}
\fontsize{9pt}{10pt}\selectfont
\begin{tabular*}{1\linewidth}{@{\extracolsep{\fill}}ccccc}
\shline
 Proj. & Imag. & PSNR $\uparrow$ & SSIM(\%) $\uparrow$ & RMSE $\downarrow$\\
\midrule
\multicolumn{2}{c}{FBP} 
& 33.59$\pm$2.69 & 76.03$\pm$10.5 & 43.78$\pm$13.8\\

\methodproj & / 
& 41.75$\pm$1.57 & 96.40$\pm$1.33 & 16.55$\pm$2.88\\

/ & SGDiff
& 40.19$\pm$1.47 & 94.49$\pm$2.17 & 19.79$\pm$3.31\\

SPDiff & SGDiff 
& 41.19$\pm$1.59 & 95.50$\pm$1.76 & 17.67$\pm$3.22\\

\methodproj & \methodimg 
& \textbf{42.82$\pm$1.05} & \textbf{96.75$\pm$0.79} & \textbf{14.51$\pm$1.69}\\
\shline
\end{tabular*}
\end{table}

\begin{figure*}[!t]
\centerline{
\includegraphics[width=1\linewidth]{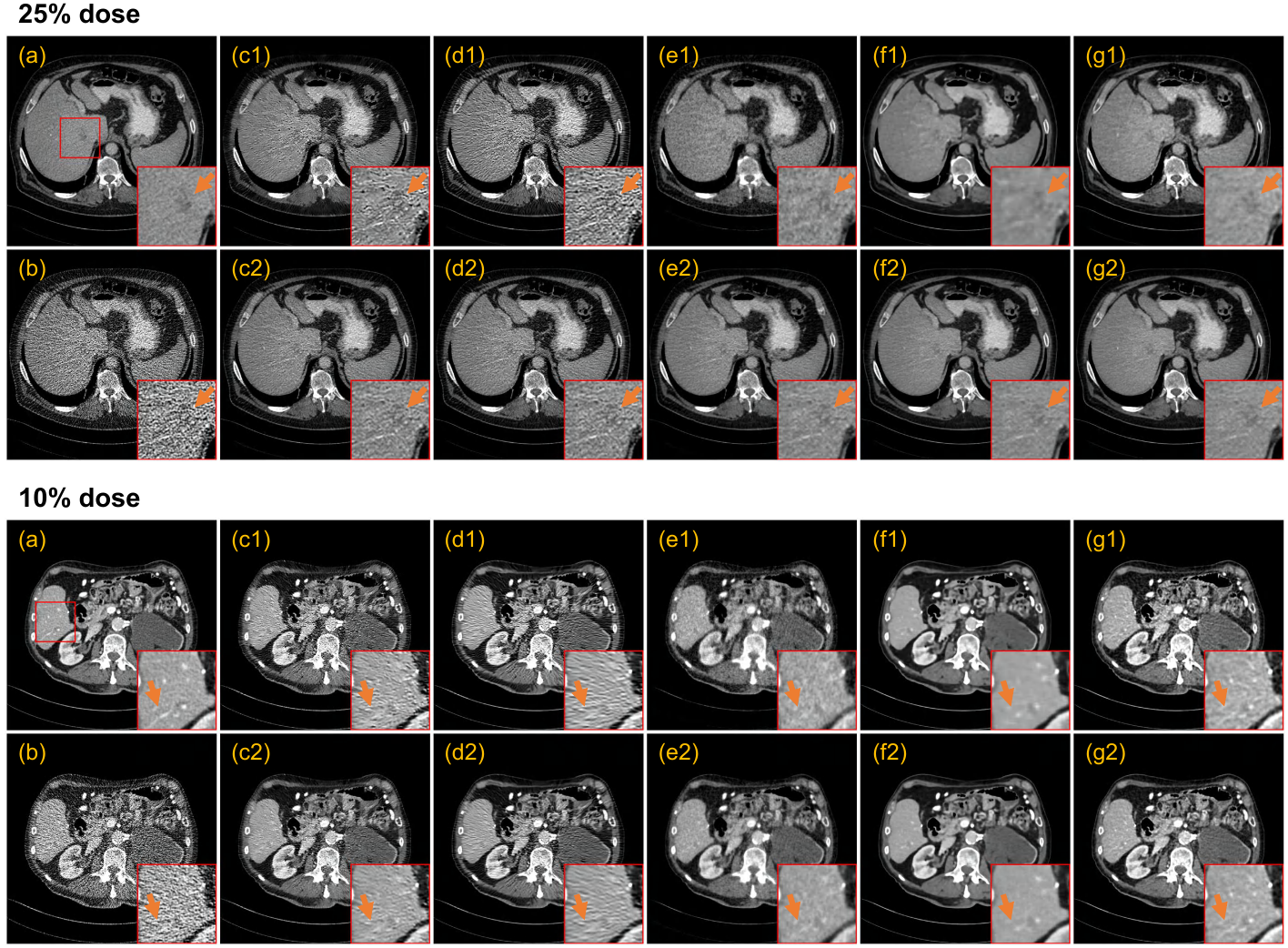}}
\caption{Qualitative results of \methodimg on the Mayo 2016 dataset using initial reconstructions $\hat{\vct{x}}_0$ provided by different algorithms. (a) NDCT image (Ground truth), (b) FBP, (c) NLM, (d) BM3D, (e) PWLS, (f) Noise2Sim, and (g) \methodproj. (c1) to (g1) show initial reconstructions from the original algorithms, while (c2) to (g2) present the final images further refined by \methodimg. The display window is [-160, 240] HU. The red ROI is zoomed in for visual comparison and the orange arrow points to a key detail.}
\label{fig:mayo2016_initinal_reconstruction}
\end{figure*}

\begin{table*}[!t]
\centering
\caption{Quantitative results of \methodimg on the Mayo 2016 dataset using initial reconstructions $\hat{\vct{x}}_0$ provided by different algorithms. The best results are highlighted in \textbf{bold}.}
\label{tab:initinal_reconstruction}
\setlength{\tabcolsep}{3pt}
\fontsize{8.5pt}{10pt}\selectfont
\begin{tabular*}{1\linewidth}{@{\extracolsep{\fill}}l|ccc|ccc}
\shline
\multirow{2}{*}{\textbf{Methods}}
& \multicolumn{3}{c|}{\textbf{25\% dose}}  
& \multicolumn{3}{c}{\textbf{10\% dose}}\\
& PSNR $\uparrow$ & SSIM(\%) $\uparrow$ & RMSE $\downarrow$
& PSNR $\uparrow$ & SSIM(\%) $\uparrow$ & RMSE $\downarrow$\\
\hline

FBP
& 33.59 & 76.03 & 43.78
& 29.40 & 58.84 & 71.17\\

\hline
NLM\hspace{7.37mm}$\rightarrow$\hspace{7.43mm}NLM+\methodimg
& 37.55~$\rightarrow$~40.54 & 90.94~$\rightarrow$~94.45 & 27.09~$\rightarrow$~19.19
& 33.64~$\rightarrow$~38.41 & 82.57~$\rightarrow$~92.83 & 42.81~$\rightarrow$~24.32\\

BM3D\hspace{5.7mm}$\rightarrow$\hspace{5.78mm}BM3D+\methodimg
& 38.15~$\rightarrow$~40.40 & 90.26~$\rightarrow$~94.15 & 26.51~$\rightarrow$~19.62 
& 34.84~$\rightarrow$~36.57 & 82.95~$\rightarrow$~87.88 & 40.05~$\rightarrow$~31.76\\

PWLS\hspace{6.05mm}$\rightarrow$\hspace{6.08mm}PWLS+\methodimg
& 36.28~$\rightarrow$~40.71 & 92.42~$\rightarrow$~95.35 & 30.67~$\rightarrow$~18.57
& 34.68~$\rightarrow$~36.65 & 88.55~$\rightarrow$~94.11 & 36.92~$\rightarrow$~29.45\\

Noise2Sim~$\rightarrow$~Noise2Sim+\methodimg
& 41.65~$\rightarrow$~42.05 & 96.12~$\rightarrow$~96.24 & 16.65~$\rightarrow$~15.86
& \textbf{40.08}~$\rightarrow$~40.77 & 95.15~$\rightarrow$~95.54 & \textbf{19.97}~$\rightarrow$~18.39\\

\hline
\methodproj\hspace{5.2mm}$\rightarrow$\hspace{8.95mm}\methodname (\textbf{ours})  
& \textbf{41.75}~$\rightarrow$~\textbf{42.82} & \textbf{96.40}~$\rightarrow$~\textbf{96.75} & \textbf{16.55}~$\rightarrow$~\textbf{14.51}
& 39.45~$\rightarrow$~\textbf{40.90} & \textbf{95.24}~$\rightarrow$~\textbf{96.07} & 21.83~$\rightarrow$~\textbf{18.35}\\
\shline 
\end{tabular*}
\end{table*}

\subsection{Model Analysis}\label{sec:Ablation_Study}
We further analyze the model by examining the benefits of the noise-aligned \methodproj and the time step matching strategy, and conduct ablation studies on the proposed modules. The models utilized for analysis are trained on the Mayo 2016 dataset.

\noindent\textbf{Analysis of \methodproj}\quad Fig.~\ref{fig:timestep_matching} shows the diffusion process of \methodproj and a sampling example of a 12.5\% dose image. Since the pre-log projection data encompasses a wide signal range $[0, 1.5]$ alongside a relatively small noise scale $[0, 0.03]$, and the predominant noise originates from high-signal background regions, subtle noise variations in low-signal areas are difficult to discern visually. Therefore, we provide the corresponding reconstructed images to facilitate clearer visualization.
First, by introducing the projection data noise model to construct the degradation operator, \methodproj achieves improved alignment between the diffusion process and the degradation process observed in LDCT images.
Second, we employ the time step matching strategy in  Eq.~\eqref{eq:shifted_Poisson_diffusion_best_t1} to determine the $\lambda_t$ closest to $I_\mathrm{ld}$ (12.5\% dose). Through visual comparison and examination of the perceptual loss curve, it is evident that the example CT image is closest to the intermediate image at $t=8$ in terms of noise levels. In this way, we can not only control the denoising level for better generalization by altering the sampling starting point but also reduce the number of sampling steps.
Finally, during the sampling process, \methodproj decomposes the denoising task to remove the noise and artifacts of the LDCT image gradually. This step-by-step method reduces the denoising difficulty at each step, ultimately enhancing reconstruction performance.

\noindent\textbf{Analysis of time step matching strategy}\quad
In our previous analysis, we highlight two advantages of the proposed time step matching strategy compared to the full sampling process: 1) it adapts the sampling process to the varying noise levels present in the test LDCT data and 2) it accelerates the inference process of the proposed \methodname. As additional evidence for these points, we conduct an experiment to investigate the performance of our method starting sampling from different time steps. Notably, 10\% dose represents the lowest dose level considered in this experiment; thus even when using the time step matching strategy, SPDiff still begins sampling from $t=T$. Consequently, \methodproj ($\tau$) and \methodproj ($T$) yield identical results at this dose level. To ensure fairness when comparing \methodname ($t_\mathrm{prj}^{*}+t_\mathrm{img}^{*}$) and \methodname ($t_\mathrm{prj}^{*}+T$), both are guided by the initial reconstruction from \methodproj ($t_\mathrm{prj}^{*}$); the sole difference lies in the starting point for image domain sampling. Fig.~\ref{fig:mayo2016_all_t} presents the reconstructed results for two representative slices. At 25\% dose, it can be observed that \methodproj ($T$) produces brighter reconstructed pixels in the ROI and a noticeable CT value drift in the residual map. Meanwhile, \methodname ($t_\mathrm{prj}^{*}+t_\mathrm{img}^{*}$) better suppresses CT value drift in the region indicated by the orange arrow compared to \methodname ($t_\mathrm{prj}^{*}+T$), while preserving more details in the chest ROI. Tab.~\ref{tab:mayo2016_all_t} presents quantitative results and inference times of different methods, showing that \methodname ($t_\mathrm{prj}^{*}+t_\mathrm{img}^{*}$) not only achieves superior quantitative metrics compared to \methodname ($t_\mathrm{prj}^{*}+T$) but also significantly reduces the sampling time required per slice.

\noindent\textbf{Ablation on different modules}\quad We first perform ablation studies to examine the contribution of each module within the proposed \methodname, utilizing the 25\% dose test data from the Mayo 2016 dataset for analysis. 
To demonstrate the effectiveness of double guidance, we also explored the performance of the model with single image guidance; the resultant model is referred to as the singly guided diffusion model (SGDiff).
When SGDiff is used alone, it employs the LDCT image as guidance. In combination with \methodproj, it relies solely on the reconstructed image as guidance. Fig.~\ref{fig:ablation_25dose} presents the qualitative results of ablation studies. All modules of our \methodname demonstrate effectiveness in reducing image noise. Although the \methodproj preserves essential details, it blurs the reconstructed image and introduces secondary artifacts. SGDiff yields sharpness results, but inevitably transfers the noise in the guidance image to the denoising image, leading to suboptimal noise elimination. SPDiff+SGDiff enhances sharpness compared to \methodproj. However, as presented in the red ROI, the lack of LDCT image guidance still limits its ability to accurately restore the key detail. Finally, \methodname, incorporating all modules, delivers the most favorable visual result. Table~\ref{tab:quantitative_results_on_ablation_25dose} presents the quantitative performance of various modules. 
Our \methodname achieves an improvement of +2.63 dB in PSNR, +2.39\% in SSIM, and -5.28 HU in RMSE compared to SGDiff.

To further establish the necessity of \methodproj and the advantages of our dual-domain framework, we investigate the performance of \methodimg across different initial reconstructions by replacing \methodproj with traditional image denoising methods NLM~\citep{buades2005non} and BM3D~\citep{dabov2007image}, an iterative reconstruction method PWLS, and Noise2Sim. These alternative methods are selected because they do not require paired ND/LDCT data for training, ensuring fair comparison with our method. We choose Noise2Sim because it demonstrates superior performance among all competing methods in our experiments, excluding Noise2Noise that is not strictly considered a self-supervised method.
Fig.~\ref{fig:mayo2016_initinal_reconstruction} presents the qualitative results of using initial reconstructions from different methods. For all methods, the proposed \methodimg further enhances the quality of the initial reconstructions and reduces quality variations between them, yielding more consistent final reconstructions. However, although \methodimg exhibits robustness, the quality of initial reconstruction still markedly influences the final outcome. DL-based methods, due to their effective suppression of noise and artifacts in the initial reconstruction, produce final reconstructed images with superior quality compared to NLM, BM3D, and PWLS. Between the DL-based methods, 
it can be observed that Noise2Sim's initial reconstructions suffer from detail loss due to over-smoothing. This introduces biases in prior information localization, challenging the subsequent processing by the \methodimg. In contrast, our \methodproj effectively retains these details, meaning the subsequent image domain refinement primarily focuses on suppressing blur and secondary artifacts.

The quantitative results in Table~\ref{tab:initinal_reconstruction} further corroborate these observations. Notably, we find that \methodimg yields better relative performance gains when applied to poorer initial reconstructions, demonstrating its robustness to different $\hat{\vct{x}}_0$. Furthermore, comparing PWLS with NLM and BM3D, as well as \methodproj with Noise2Sim, we observe that \methodimg achieves greater performance improvements on PWLS and \methodproj, despite similar quantitative performance in their initial reconstructions. We attribute this to both PWLS and our \methodproj leveraging dual domain complementary information (projection and image), providing richer details in $\hat{\vct{x}}_0$. This enhances the subsequent refinement by \methodimg, which primarily models the NDCT image distribution and therefore gains less complementary prior information from purely image domain methods.
\section{Discussion}
\label{sec:dicussion}
Inspired by the noise characteristics in different domains, the introduced dual-domain \methodname takes advantage of CT physics for modeling and achieves better reconstruction and generalization. 
We also highlight that the training of \methodname only requires NDCT data. Compared to the supervised training that requires strictly paired ND/LDCT data, our \methodname greatly reduces the difficulty of data collection.
Compared to previous self-supervised models that adapt to the LDCT data at specific doses, our model directly models the distribution of NDCT data, making it dose-agnostic. Coupled with the proposed time step matching strategy, our method exhibits superior generalization performance across multiple dose levels. In the following subsections, we elaborate on the benefits of utilizing the number of X-ray incident photons $I_0$, distinguish our \methodname from related diffusion-based works, highlight advantages over simulation-based approaches, and discuss limitations of our method.

\subsection{Benefits of the \methodproj}
In our \methodname, \methodproj serves as a crucial component by providing a high-quality initial reconstruction for subsequent refinement. Here, we summarize the key contributions of \methodproj to elaborate its distinct advantages and necessity within our framework. 1) Fig.~\ref{fig:mayo2016_initinal_reconstruction} shows that the quality of initial reconstruction significantly impacts the final outcome. Traditional methods suffer from insufficient noise and artifact suppression, while deep learning algorithms lose details through over-smoothing. The loss of details is undesirable for \methodimg as it leads to inaccurate prior information localization and reduced credibility of generated details. The proposed \methodproj provides high-quality reconstruction results even without image domain refinement, comparable to advanced self-supervised methods while preserving more details, though it exhibits some blurring due to limitations of projection data processing. 2) Image domain methods can serve as an alternative for initial reconstruction when projection data is unavailable. However, as Table~\ref{tab:initinal_reconstruction} demonstrates, the complementary information derived from the projection domain enables our image domain model \methodimg to achieve superior reconstruction quality. This interchangeability demonstrates our method's flexibility without diminishing the contributions of \methodproj. 3) \methodproj adaptively selects the sampling start point via a timestep matching strategy based on the number of incident photons $I_0$, ensuring robust initial reconstructions across diverse dose levels. In contrast, traditional image domain methods or iterative reconstruction algorithms require dose-specific manual hyperparameter adjustments, while self-supervised methods need additional training for varying dose settings.

\subsection{Benefits of Utilizing $I_0$}
Our \methodname leverages the number of X-ray incident photons $I_0$, which directly reflects the noise level in the input pre-log projection, for selecting the sampling start point. Given that compared methods do not utilize such physical parameters, we further clarify the fairness and feasibility of our experiments, while highlighting the advantages of incorporating $I_0$ compared to other generalizable LDCT reconstruction works.
Our \methodname leverages the number of X-ray incident photons $I_0$, which directly reflects the noise level in the input pre-log projection, for selecting the sampling start point. Since comparative methods do not utilize such physical parameters, we further clarify the fairness and feasibility of our experiments while highlighting the advantages of incorporating $I_0$ compared to other generalizable LDCT reconstruction works.

\noindent\textbf{Fairness of comparison}\quad 
We emphasize that leveraging $I_0$ for adaptive reconstruction is a core innovation of our method, intrinsically designed to improve generalization, rather than an extrinsic advantage introduced solely for evaluation at any specific dose level.
In contrast, competing methods generally lack mechanisms for noise adaptation based on physical parameters. 
We have implemented all competing methods faithfully according to their original papers or official code, utilizing all information available within their respective frameworks. 
Therefore, our comparison rigorously evaluates the inherent generalization capability of each method on seen and unseen dose levels within the confines of its original design, ensuring a fair evaluation. Although adding a noise level to existing methods is feasible through various approaches, it has not been well benchmarked or sufficiently validated in their original works. We hope that our study can serve as a benchmark for the research community.

\noindent\textbf{Clinical feasibility}\quad While our method requires $I_0$ to determine an appropriate time step $t_\mathrm{prj}^{*}$ as the starting point for sampling, obtaining this parameter is practical in clinical settings. $I_0$ can typically be estimated either from air calibration scans (routinely provided by manufacturers) or by analyzing unattenuated regions in raw measurements where detectors receive X-rays without obstruction~\citep{hsieh2003computed, Thorsten2003computed}. Therefore, the requirement of $I_0$ poses no significant technical barrier to clinical implementation of our method.

\noindent\textbf{Comparison with other generalizable reconstruction works}\quad We further analyze the advantages of \methodname over recent representative generalizable reconstruction works like PDF~\citep{xia2021ct} and CoreDiff~\citep{gao2024corediff}. Both PDF and CoreDiff require paired ND/LDCT data for supervised training. PDF additionally requiring dose information paired with training data, while \methodname operates without these paired data constraints. During inference, PDF needs precise dose information matching training settings, while CoreDiff requires fine-tuning using a single LDCT image (un)paired with NDCT from the test data. Therefore, the $I_0$ required by our method is generally easier to acquire compared to these requirements.

\subsection{Differences with Related Diffusion-based Works}
There are some recent studies employing diffusion models for image restoration or low-dose CT reconstruction tasks. We clarify the distinctions between these approaches and our \methodname to further elucidate the contributions of this work.

\noindent\textbf{Comparison of \methodproj with recent works}\quad
The proposed \methodproj is a customized generalized diffusion model to denoise LDCT projection by aligning the noise model in the pre-log projection domain. To the best of our knowledge, this is the first work to incorporate an explicit noise model within a generalized diffusion model for LDCT reconstruction. Here, we discuss the advantages of \methodproj compared to recent related works:
1) Existing generalized diffusion-based LDCT reconstruction methods construct generalized diffusion processes in the image domain using paired ND/LDCT images to implicitly model noise distribution information~\citep{gao2024corediff, lu2024pridediff}. Different from them, our \methodproj effectively models the noise in pre-log projection domain through an explicit shifted Poisson diffusion process, requiring only clean NDCT data for training. 2) While the self-supervised method DDM$^2$~\citep{xiang2023ddm} can be trained using only degradation images, it employs the pretrained diffusion model as a Gaussian noise remover, which is suboptimal for LDCT reconstruction since CT noise distribution does not conform to the Gaussian distribution in both projection and image domains. 3) Unsupervised diffusion-based methods use diffusion models pretrained on clean data as generative priors, but GDP and Dn-Dp require time-consuming sampling from random Gaussian noise, while DR2 requires manual selection of sampling starting points and truncation timesteps for different degradation levels. Our SPDiff can apply the proposed timestep matching strategy to adaptively select sampling starting points based on the number of incident photons $I_0$, which not only adapts test data at various dose levels but also accelerates sampling, requiring only about 25\% of DR2's inference time (the fastest comparison method) when sampling a $672 \times 672$ projection data.

\noindent\textbf{Comparison of \methodimg with recent works}\quad
The main improvement of the proposed \methodimg, lies in its doubly guidance mechanism, which enhances the localization accuracy of prior information. We outline the advantages compared to recent works as follows. 1) \emph{Compared to DR2}: DR2 uses only the degraded image for guidance, requiring manual selection of a suitable sampling starting point to fit the input noise levels and specific band-pass filters to mitigate degradation effects for images with different noise levels. In contrast, our \methodimg can adaptively select the sampling starting point from the initial reconstruction, and its doubly guidance mechanism inherently reduces the negative impacts of noises and artifacts in LDCT images without needing manually selected filters, achieving better results with reduced inference times.
2) \emph{Compared to GDP}: GDP employs two guidance strategies, named GDP-$\vct{x}_0$ and GDP-$\vct{x}_t$. 
Both necessitate a known degradation model to match the guided image and the sampling image, preventing misalignment errors. Unfortunately, its simple degradation assumptions are insufficient for complex LDCT image noise.
In contrast, our \methodimg guides the clean estimate $\tilde{\vct{x}}_0$ using a high-quality initial reconstruction, eliminating the need for an explicit CT image degradation model. Furthermore, it innovatively guides the noisy sample $\vct{x}_t$ using the LDCT image, incorporating valuable information from the degraded image while effectively mitigating its adverse effects.
3) \emph{Compared to Dn-Dp}: While Dn-Dp performs image domain guidance using the LDCT image and balances data fidelity and noise suppression via adaptive weights---a technique also present in \methodimg---a key advantage of our \methodimg is the incorporation of additional information from the initial reconstruction. As evidenced by results in Fig.~\ref{fig:mayo2016_initinal_reconstruction} and Table~\ref{tab:initinal_reconstruction}, utilizing complementary information derived from both the projection data (provided by the initial reconstruction) and the LDCT image allows \methodimg to achieve more precise prior localization and provide improved reconstruction results compared to methods relying only on image domain guidance.

\subsection{Benefits Over Simulation-Based Approaches}
Clinical CT scanners typically utilize standardized protocols that can be predetermined and are well-established, making it feasible to simulate CT images at specific dose levels from NDCT data for supervised training. However, our proposed method offers several key advantages over these simulation-based methods:

\noindent\textbf{Avoiding simulation biases}\quad
Simulation-based methods often suffer from a distribution shift between simulated and actual LDCT data, potentially impairing performance in real clinical settings. Additional fine-tuning is typically needed to narrow this disparity. Our method circumvents these issues by training solely on more readily available NDCT data from real-world scenarios (such as large-scale health screening programs), reducing data acquisition costs and avoiding biases from simulation or non-clinical training data.

\noindent\textbf{Building a fundamental model for clinical evaluation}\quad 
In clinical practice, CT scanning doses vary widely, with actual tube currents differing across manufacturers even at identical nominal levels due to proprietary protocol implementations. Training specific models for such diverse dose scenarios presents significant challenges, while developing and deploying individual models for each dose level substantially increases computational burden. In contrast, our \methodproj aligns to the ``Poisson + Gaussian'' noise model in pre-log raw CT measurements---stemming from physics characteristics---and performs adaptive reconstruction based on the number of X-ray photon counts. Therefore, our \methodname provides a fundamental solution for clinical LDCT that adapts to various dose scenarios and can be fine-tuned for cross-device and cross-center applications.

\noindent\textbf{Benefits for new protocols exploration}\quad 
While standardized protocols meet the imaging requirements for most clinical scenarios, equipment manufacturers, radiologists, and researchers are actively dedicated to exploring improved scanning protocols, particularly for pediatric patients and specific clinical challenges such as low-dose lung cancer screening and follow-up examinations for patients. However, simulation-based, protocol-specific methods yield suboptimal results when applied to these new protocols. In contrast, our dose-agnostic training paradigm enables the proposed method to seamlessly adapt to varying dose levels from new protocols without requiring retraining, facilitating research and development of novel scanning techniques.

\subsection{Limitations}
Here, we acknowledge some limitations in our \methodname. First, we utilized the shifted Poisson noise model to approximate the noise model in pre-log LDCT projections, considering a crucial factor---the number of X-ray incident photons. In clinical scanning scenarios, the noise level in pre-log LDCT projections is influenced by various factors like tube voltage, filters, and patient differences. Our future work will consider these factors to improve the diffusion process, making it more consistent with the practical degradation process of pre-log LDCT projection data. Second, while our fan-beam simulation captures a key element of the reconstruction pipeline---many commercial CT reconstruction workflows rebin helical CT projections into 2D fan-beam projections before backprojection~\citep{hoffman2016freect_wfbp,gong2021deep}---such simulated projection data cannot fully replicate the complex physical phenomena present in real measured projection data, including scatter, detector non-linearities, and other artifacts. Moving forward, we aim to address this limitation by collecting real pre-log CT projections to further validate and enhance our method for processing authentic clinical CT pre-log projection data. Third, although the proposed method demonstrates good generalization performance on unseen doses and unseen datasets with different noise distributions and scanning protocols, clinical CT data presents a more complex and diverse noise pattern. We intend to conduct more comprehensive experiments across multi-center and multi-device datasets to thoroughly evaluate the generalization capabilities of our method in varied real-world clinical settings.
Fourth, while our \methodname has less inference time than single-domain diffusion-based methods, image super-resolution is the main time-consuming step in sampling $512\times512$ high-resolution CT images, with each slice taking over 0.5s. We plan to explore other ordinary differential equation samplers and perform refinement in the latent space to accelerate sampling in high-resolution CT image sampling.
Fifth, although the ablation study in Fig.~\ref{fig:ablation_25dose} shows that \methodname provides clearer images with better detail preservation compared to \methodproj, this improvement comes at the cost of increased computational complexity. Future work will explore parameter-efficient techniques such as latent space sampling and knowledge distillation. Additionally, while the noise model for CT image domain remains unknown, some studies approximate it using Gaussian mixture models or spatially correlated noise models~\citep{li2022noise, divel2019accurate}. We plan to explore specialized noise modeling methods that could potentially reduce model parameters while maintaining reconstruction quality.
\section{Conclusion}
\label{sec:conclusion}
In this work, we introduced a novel noise-inspired diffusion model (\methodname) for generalizable low-dose CT reconstruction. Our \methodname utilizes 1) a shifted Poisson diffusion model to align with the noise model for pre-log LDCT projection data denoising, 2) a {doubly guided diffusion model to locate prior information more precisely for image domain refinement},  3) a time step matching strategy to adaptively control denoising and refinement while accelerate sampling, and 4) only NDCT data for training.  Extensive experimental results validate the effectiveness and generalization of our \methodname.

\section*{Declaration of competing interest}
I have nothing to declare.

\section*{Data availability}
The authors do not have permission to share data.

\section*{Acknowledgments}
This work was supported in part by National Natural Science Foundation of China (Nos.  62471148, 62101136, and U21A6005), National Key R\&D Program of China (Nos.  2024YFA1012000 and 2024YFC2417800).
\bibliographystyle{model2-names}\biboptions{authoryear}
\bibliography{main}

\end{document}